\documentclass[aps,prd,preprint,superscriptaddress,tightenlines,nofootinbib]{revtex4}

\usepackage{rotating} 
\usepackage{graphicx} 
\usepackage{epsfig}

\newcommand{\rhoe}{\rho^-e^+\nu} 
\newcommand{\non}{\nonumber} 
\newcommand{\kste}{K^{*-}e^+\nu} 
\newcommand{\ke}{K^-e^+\nu} 
\newcommand{\pie}{\pi^-e^+\nu} 
\newcommand{\drhoe}{D^0\to \rho^-e^+\nu} 
\newcommand{\dkste}{D^0\to K^{*-}e^+\nu} 
\newcommand{\dke}{D^0\to K^-e^+\nu} 
\newcommand{\dpie}{D^0\to \pi^-e^+\nu}

\newcommand{\dkpi}{D^0\to K^-\pi^+} 
\newcommand{\dkpp}{D^0\to K^-\pi^+\pi^0} 
\newcommand{\dkppp}{D^0\to K^-\pi^+\pi^0\pi^0} 
\newcommand{\dkpppc}{D^0\to K^-\pi^+\pi^+\pi^-} 
\newcommand{\dksp}{D^0\to K_S\pi^0}
\newcommand{\dkspp}{D^0\to K_S\pi^+\pi^-}
\newcommand{\dksppp}{D^0\to K_S\pi^+\pi^-\pi^0}
\newcommand{\dkk}{D^0\to K^-K^+} 
 
\newcommand{\dppp}{D^0\to \pi^-\pi^+\pi^0}

\newcommand{\vcs}{V_{cs}} 
\newcommand{\vcd}{V_{cd}}

\newcommand{\kpi}{K^-\pi^+} 
\newcommand{\kpp}{K^-\pi^+\pi^0} 
\newcommand{\kppp}{K^-\pi^+\pi^0\pi^0} 
\newcommand{\kpppc}{K^-\pi^+\pi^+\pi^-} 
\newcommand{\ksp}{K_S\pi^0}
\newcommand{\kspp}{K_S\pi^+\pi^-}
\newcommand{\ksppp}{K_S\pi^+\pi^-\pi^0}
\newcommand{\ppp}{\pi^-\pi^+\pi^0} 
\newcommand{\kk}{K^-K^+}

\usepackage{graphicx}
\usepackage{dcolumn}
\usepackage{bm}

\begin{document}

\preprint{CLEO CONF 04-3}   
\preprint{ICHEP04 ABS8-0781}      

\title{First CLEO-c Results on Exclusive $D^{0}$ 
              Semileptonic Decays }
\thanks{Submitted to the 32$^{\rm nd}$ International Conference on High 
        Energy Physics, Aug 2004, Beijing}

\author{K.~Y.~Gao}
\author{D.~T.~Gong}
\author{Y.~Kubota}
\author{B.W.~Lang}
\author{S.~Z.~Li}
\author{R.~Poling}
\author{A.~W.~Scott}
\author{A.~Smith}
\author{C.~J.~Stepaniak}
\author{J.~Urheim}
\affiliation{University of Minnesota, Minneapolis, Minnesota 55455}
% ...
\collaboration{CLEO Collaboration} 

\noaffiliation

\date{\today}

\begin{abstract}

  Based on a data sample of 60 pb$^{-1}$ collected at the 
  $\psi(3770)$ resonance with the CLEO-c detector at CESR, 
  we present improved measurements of absolute branching 
  fractions for exclusive $D^0$ semileptonic decays 
  into $\ke$, $\pie$ and $\kste$; and the first observation
  and absolute branching fraction measurement of 
  $D^{0} \to \rhoe$.

\end{abstract}

\maketitle

\section{INTRODUCTION} 

The quark mixing parameters in the Cabibbo-Kobayashi-Maskawa 
(CKM) matrix are fundamental constants that must be
determined from experiments.
Semileptonic $D$ meson decays are of great physics interest because
they are relatively simple to handle theoretically. 
The decay matrix for semileptonic $D$ meson decay decouples into 
a weak current (describing the $W \ell \nu$ vertex), 
and a strong current (for the $W c \bar{q}$ vertex) that is 
parameterized through form-factor
functions of the invariant mass ($q^{2}$) of the $W$ exchanged.
The form factors can not be easily computed in quantum 
chromodynamics (QCD) since they are affected by significant
nonperturbative contributions. 
That is the main source of uncertainty in the extraction of the 
CKM matrix element from the simple decay processes. 
Precise experimental measurements are needed to guide 
theoretical progress in this area. 
Charm semileptonic decays allow measurements of the form factors
and CKM matrix elements $V_{cs}$ and $V_{cd}$.
Using Heavy Quark Effective Theory (HQET) or lattice gauge 
techniques, the measured charm form factors can be related to 
those needed to interpret $b \to u$ transition and the 
measurement of the CKM matrix element $V_{ub}$.

\begin{figure}[htbp] 
\epsfxsize=0.6\textwidth  
\center{\mbox{\epsffile{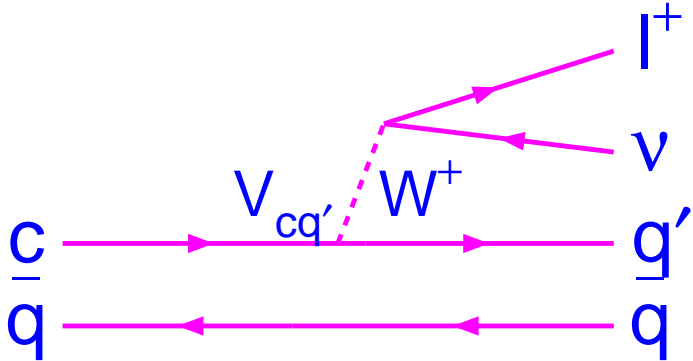}}}
\caption{Feynman diagram for charm meson semileptonic decays.}
\label{fig1} 
\end{figure}

While measurements of form factors in all charm exclusive semileptonic
decays are important, those in pseudoscalar-to-pseudoscalar
transitions are the easiest to perform.
The differential decay rate for the exclusive semileptonic decay 
$D\to P\ell\nu$ ($P$ stands for a pseudoscalar meson) with 
the electron mass effects neglected can be expressed 
as~\cite{theory}:
\begin{eqnarray}
\frac{d\Gamma}{dq^2}=\frac{G_F^2}{24\pi^3}\left|V_{cq'}\right|^2p_P^3
\left|f_+(q^2)\right|^2
\label{diffrate} 
\end{eqnarray}   
where $G_F$ is the Fermi coupling constant, $q^2$ is the 
four-momentum transfer squared between the parent $D$ meson 
and the final state meson, $p_P$ is the momentum 
of the pseudoscalar meson in the $D$ rest frame, and $V_{cq'}$ is 
the relevant CKM matrix element, either $\vcs$ or $\vcd$. 
$f_+(q^2)$ is the form factor that measures the probability that 
the flavor changed quark $q'$  and the spectator quark $\bar q$ 
in Fig.~\ref{fig1} will form a meson in the final state.  
The corresponding branching fractions can be obtained from 
\begin{eqnarray}
{\cal B}(D\to P\ell\nu)=\tau_D\times\int\limits_{q^2}dq^2
                        \frac{d\Gamma}{dq^2} 
\end{eqnarray}

\section{DETECTOR AND DATA SET}

The data sample used in this analysis was produced at the
$\psi(3770)$ resonance with the Cornell Electron Storage Ring (CESR) 
and collected with the general purpose CLEO-c 
detector~\cite{cleo-c,cleoc}.
The total integrated luminosity of this data set is 60 pb$^{-1}$.

The CLEO-c tracking system covers about 95\% of the 4$\pi$ solid
angle with a large cylindrical drift chamber and an all stereo
wire vertex chamber. 
These tracking devices also provide specific-ionization ($dE/dx$) 
measurements for particle identification.
The electromagnetic calorimeter is made of 7800 cesium iodide
crystals with a total solid-angle coverage of 95\% of 4$\pi$.
The calorimeter is crucial for electron identification and 
provides excellent efficiency and energy resolution for photons,
yielding a typical mass resolution for $\pi^{0}$ reconstruction
around 6 MeV/$c^2$ (standard deviation).
The Ring Imaging Cherenkov (RICH) detector
covers 80\% of the 4$\pi$ solid angle and provides superb particle 
identification capability.

\section{EVENT RECONSTRUCTION AND SELECTION} 

For this analysis we take full advantage of the unique kinematics 
of $D\bar D$ production at the $\psi(3770)$ resonance, just above
charm threshold.
We first select events in which a $D^{0}$ meson is fully
reconstructed in one of the nine hadronic final states:
$K^-\pi^+$,           $K^-\pi^+\pi^0$, $K^-\pi^+\pi^0\pi^0$, 
$K^-\pi^+\pi^+\pi^-$, $K_S\pi^0$,      $K_S\pi^+\pi^-$, 
$K_S\pi^+\pi^-\pi^0$, $\pi^+\pi^-\pi^0$, and $K^-K^+$.
Charge conjugation is implied here and throughout this paper.
Within these tagged events we select the subset in which the 
$\bar{D^{0}}$ meson decayed semileptonically to a specific final state.  
The efficiency-corrected ratio of the event yields gives the absolute 
branching fraction for the exclusive semileptonic decay mode.  
This branching fraction is independent of the luminosity of the 
experiment and benefits from the cancellation of many systematic 
uncertainties.  
This selection procedure provides an exceptionally clean sample of 
semileptonic decays that is ideal for determination of form-factor 
parameters and CKM elements, although our current data sample 
is not yet adequate for these determinations.

\begin{figure}[htbp]
\epsfxsize=0.33\textwidth\centerline{\mbox{
                         \epsffile{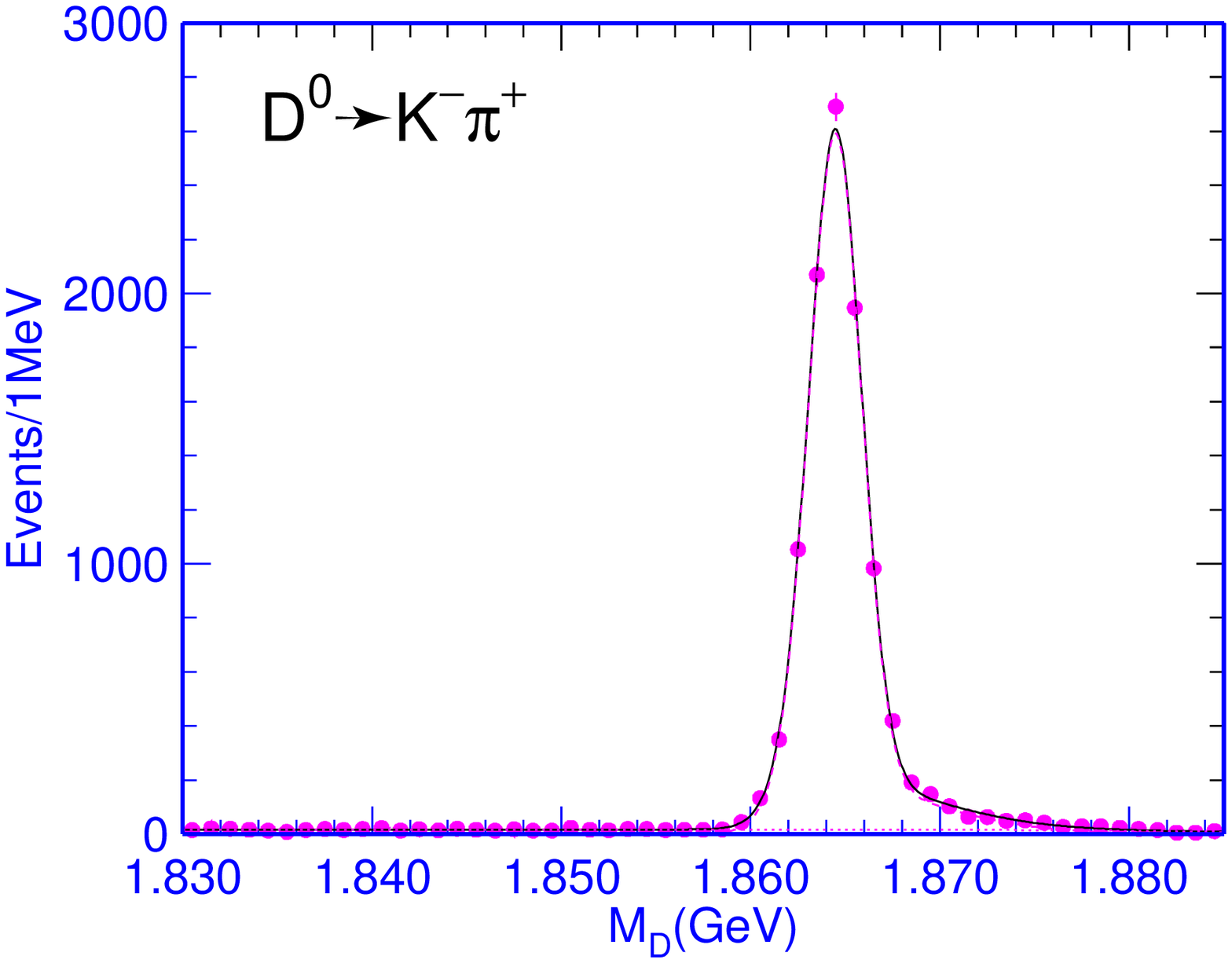} 
\epsfxsize=0.33\textwidth\epsffile{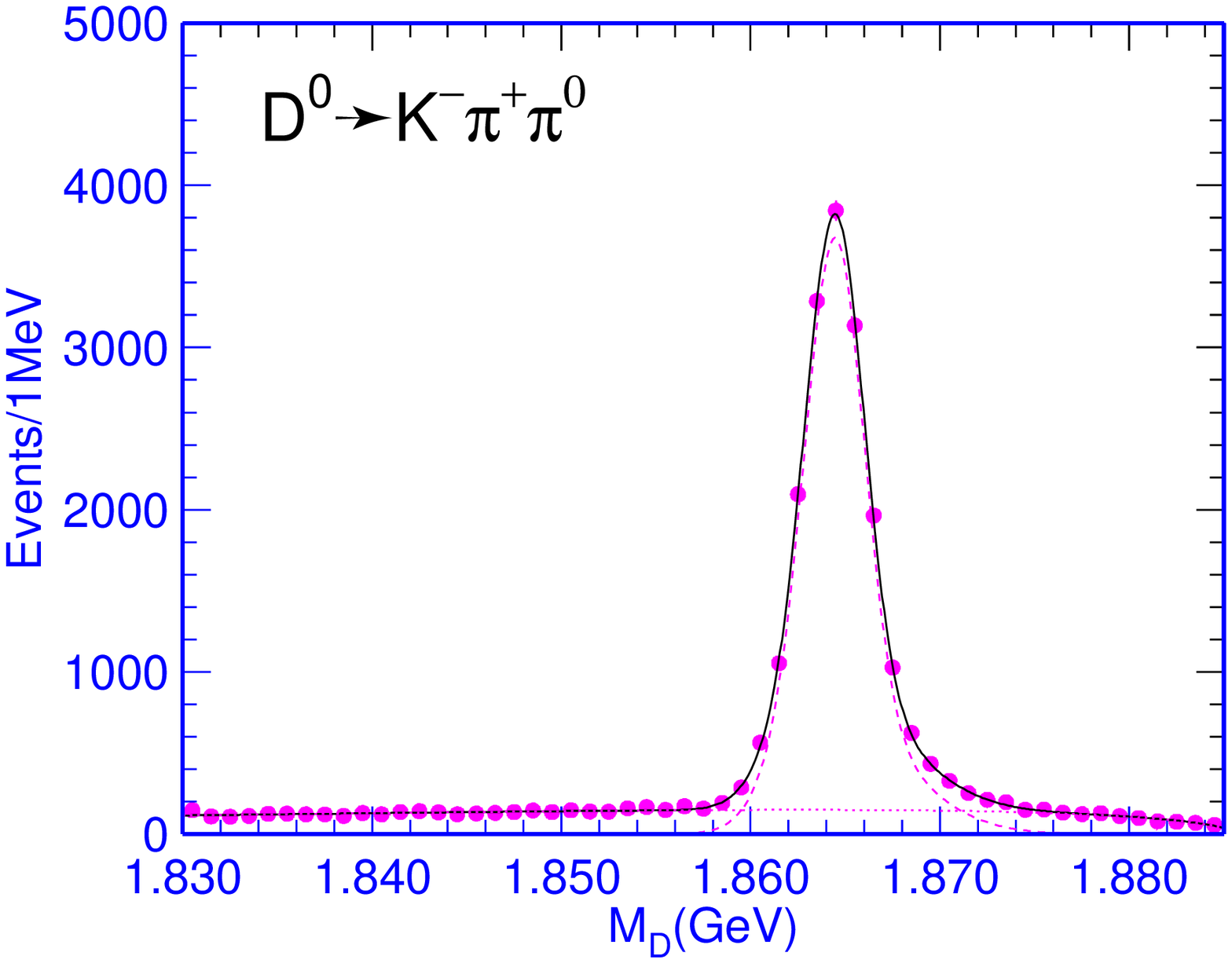}  
\epsfxsize=0.33\textwidth\epsffile{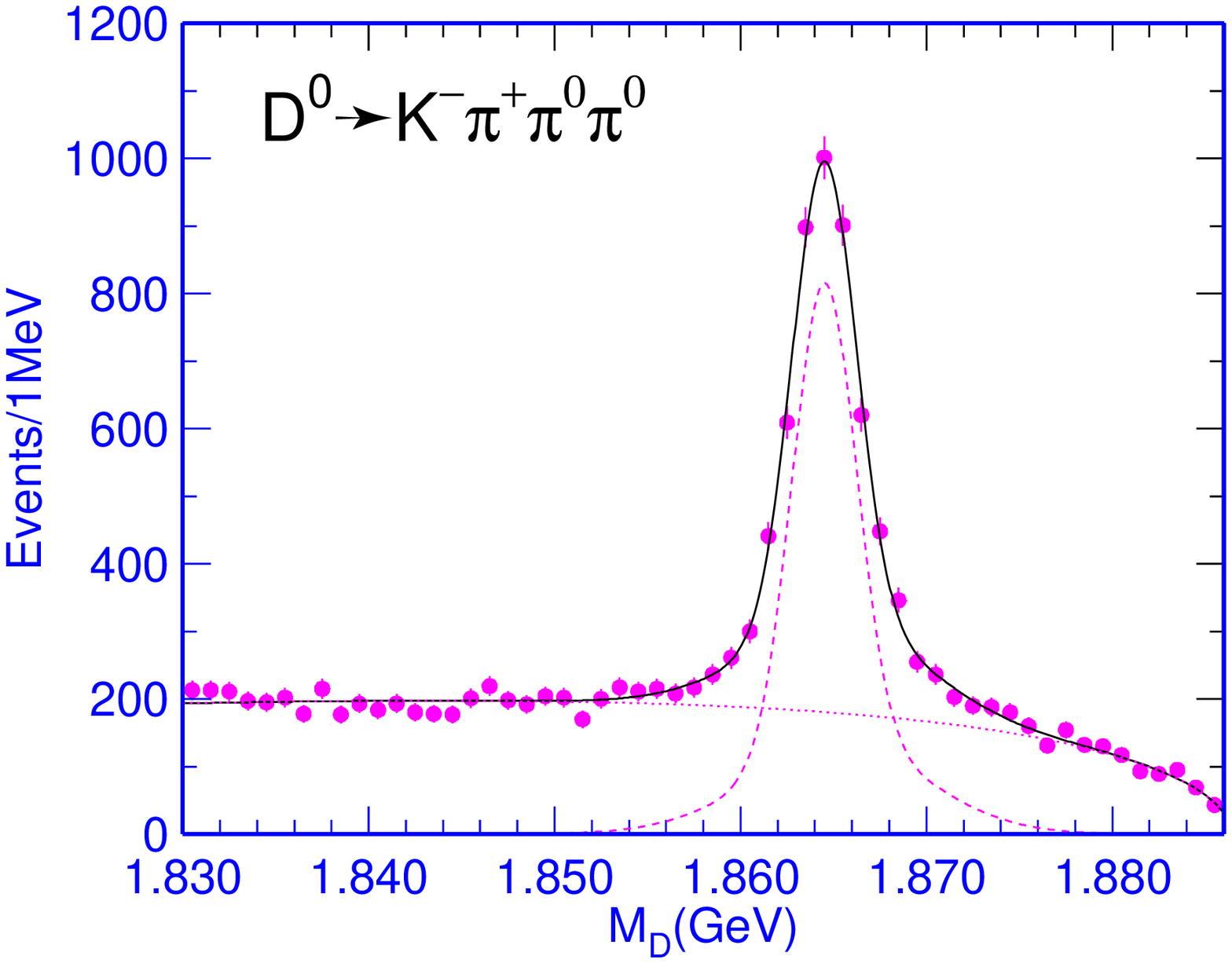}}}
\epsfxsize=0.33\textwidth\centerline{\mbox{
                         \epsffile{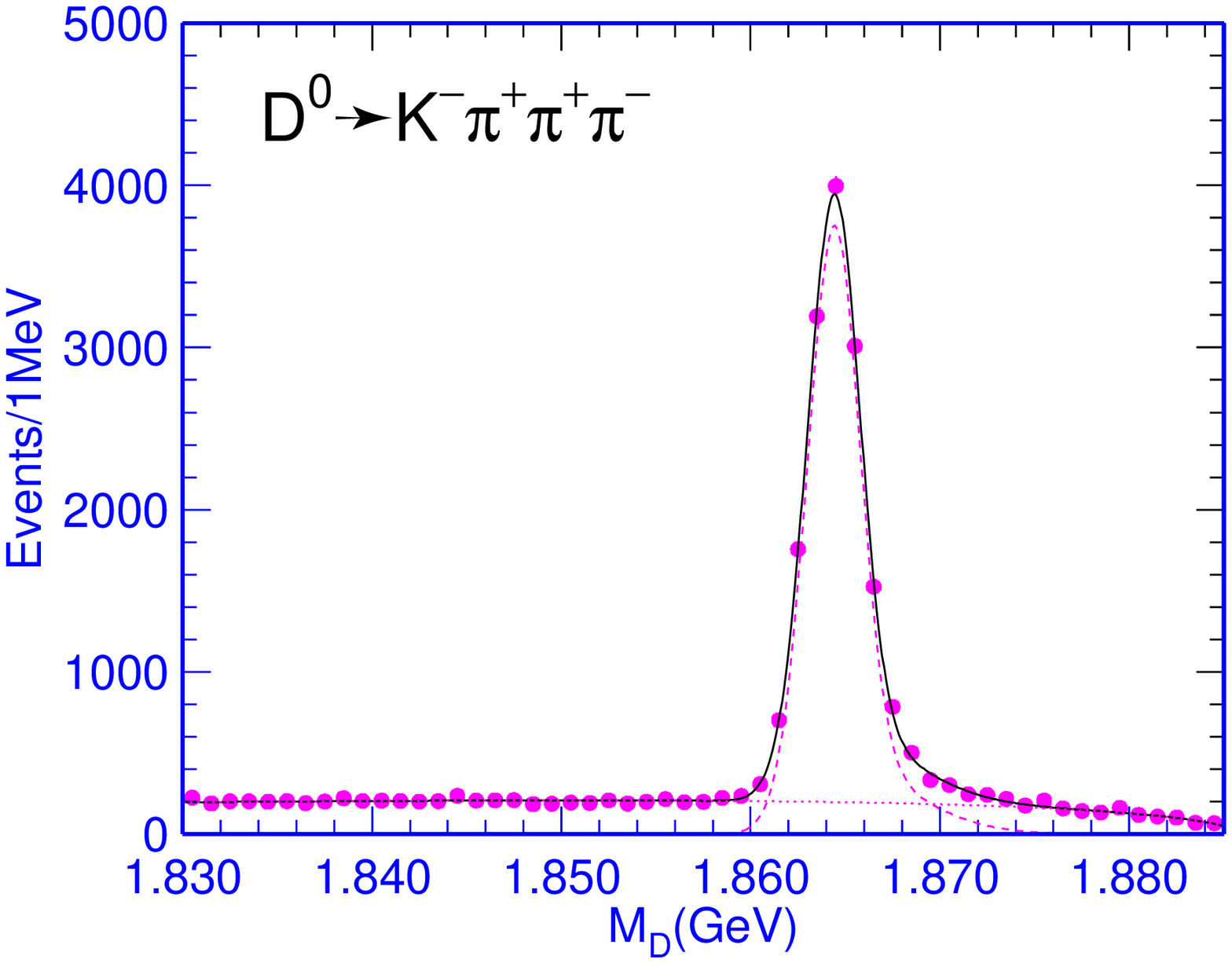} 
\epsfxsize=0.33\textwidth\epsffile{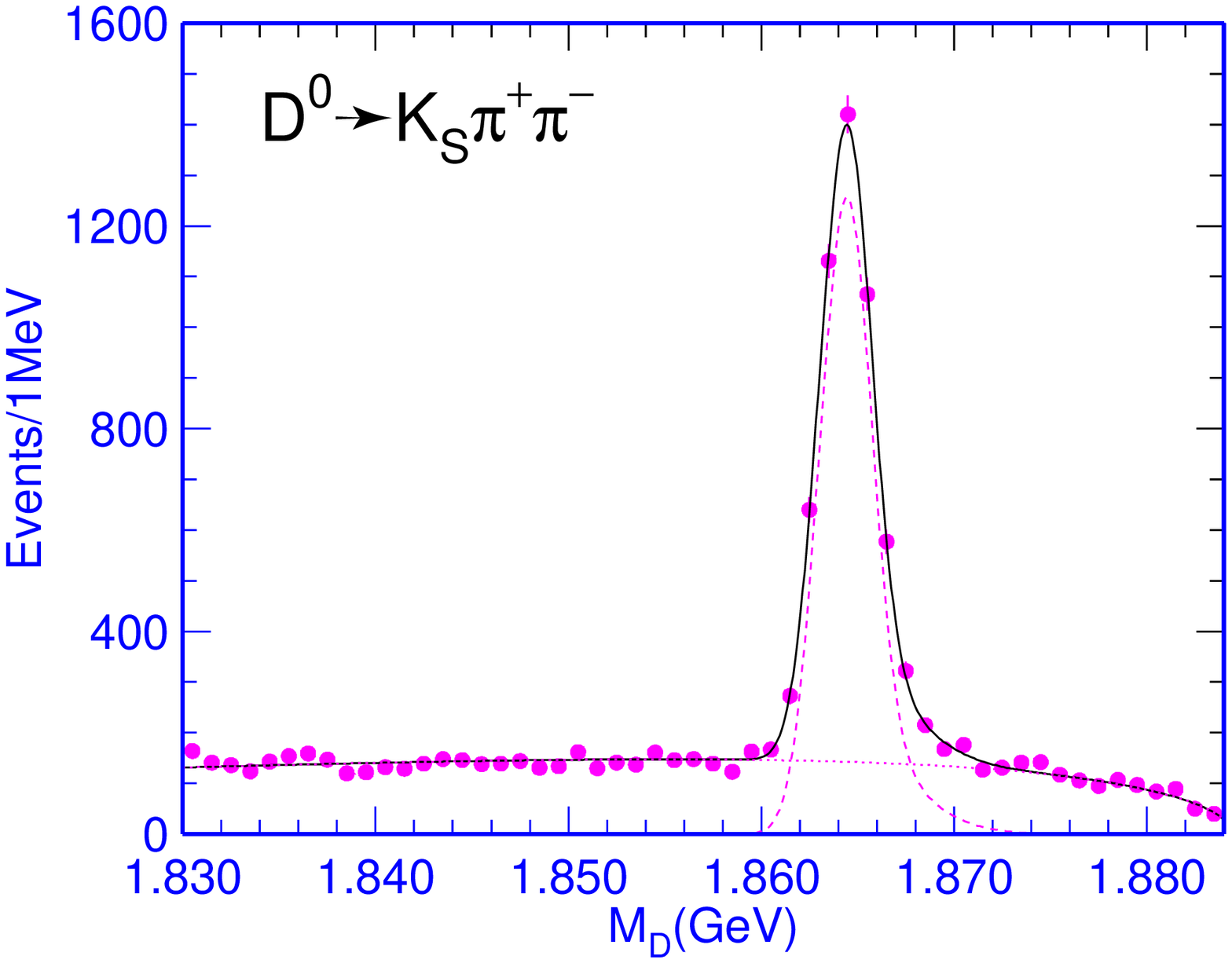}
\epsfxsize=0.33\textwidth\epsffile{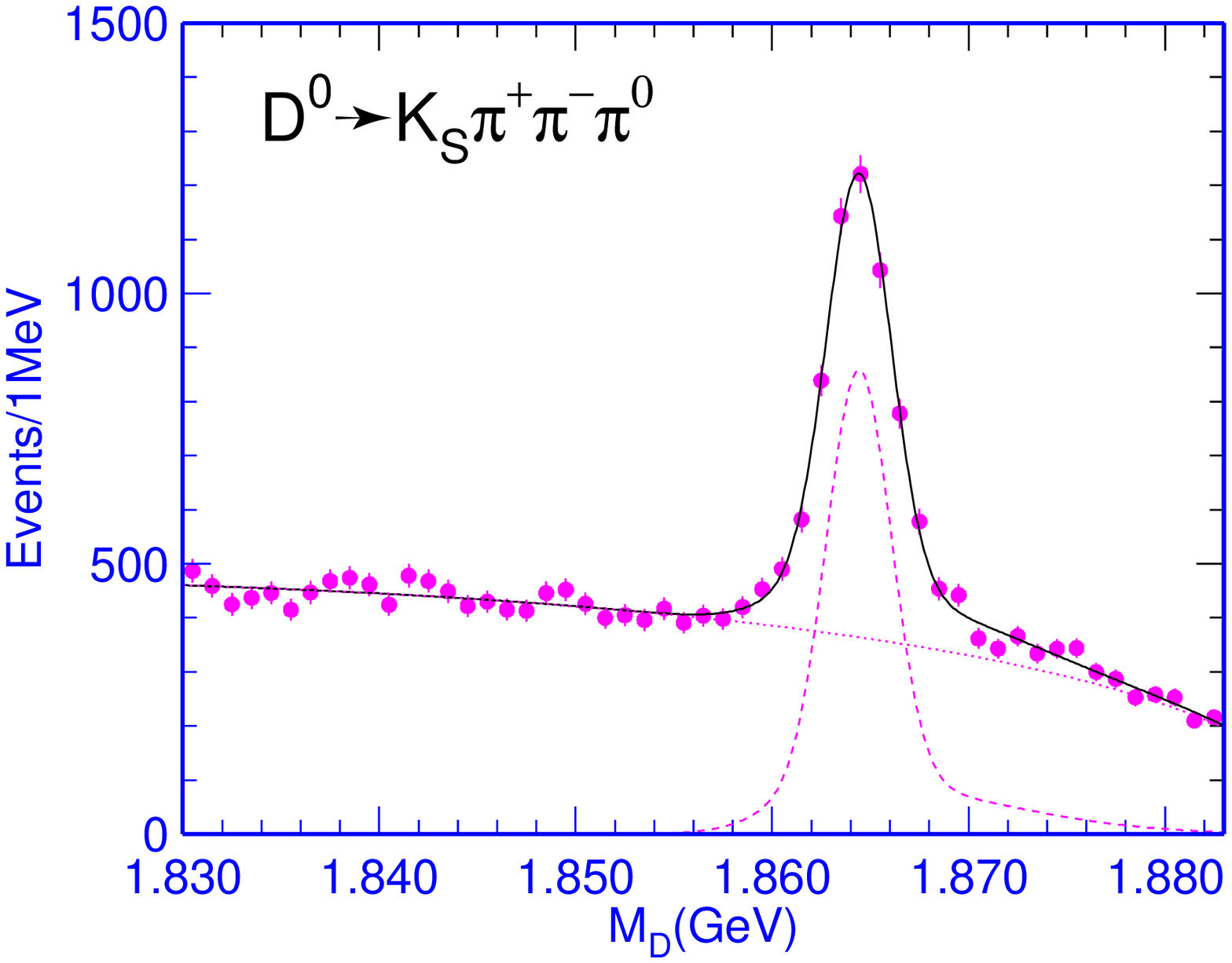}}}
\epsfxsize=0.33\textwidth\centerline{\mbox{
                         \epsffile{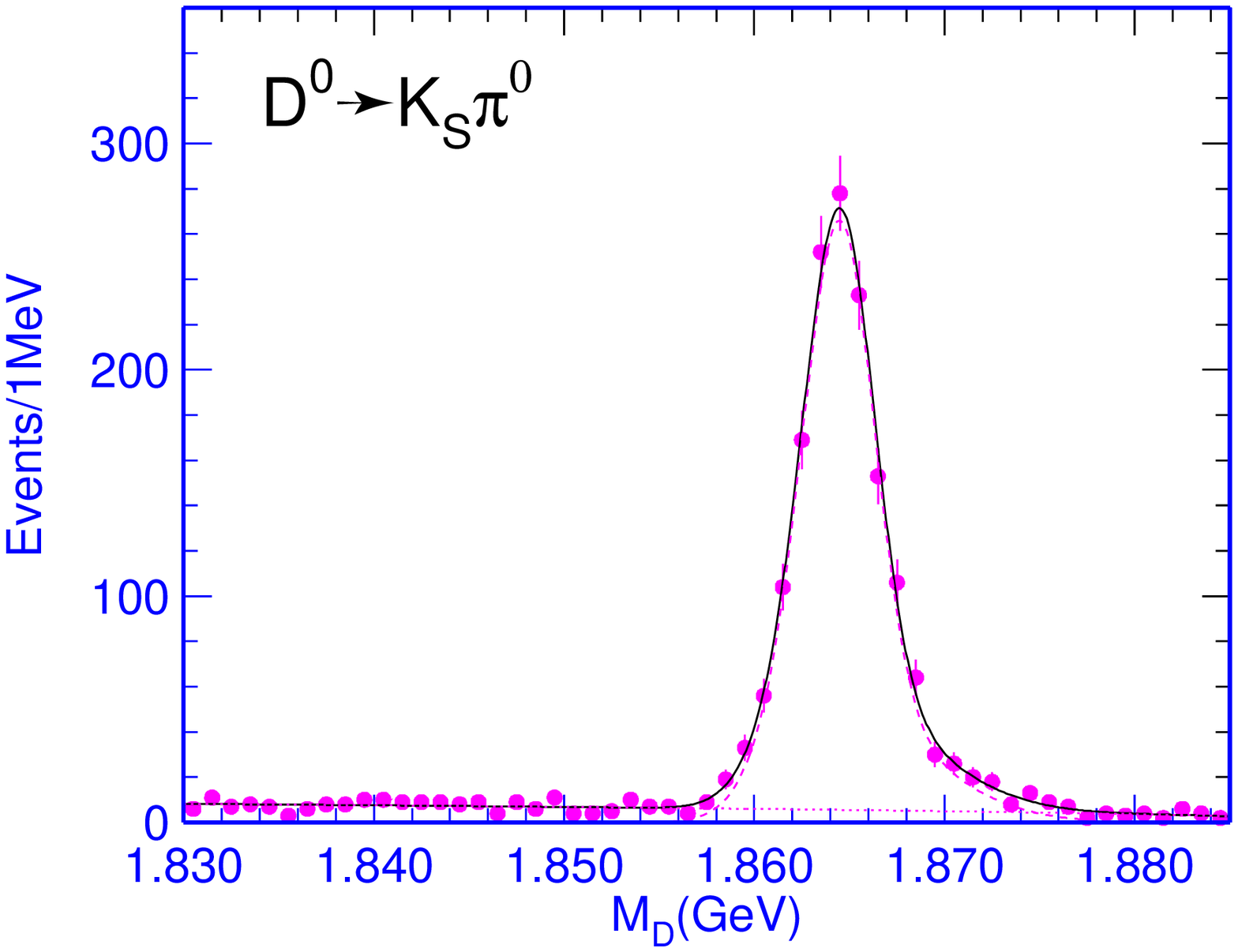} 
\epsfxsize=0.33\textwidth\epsffile{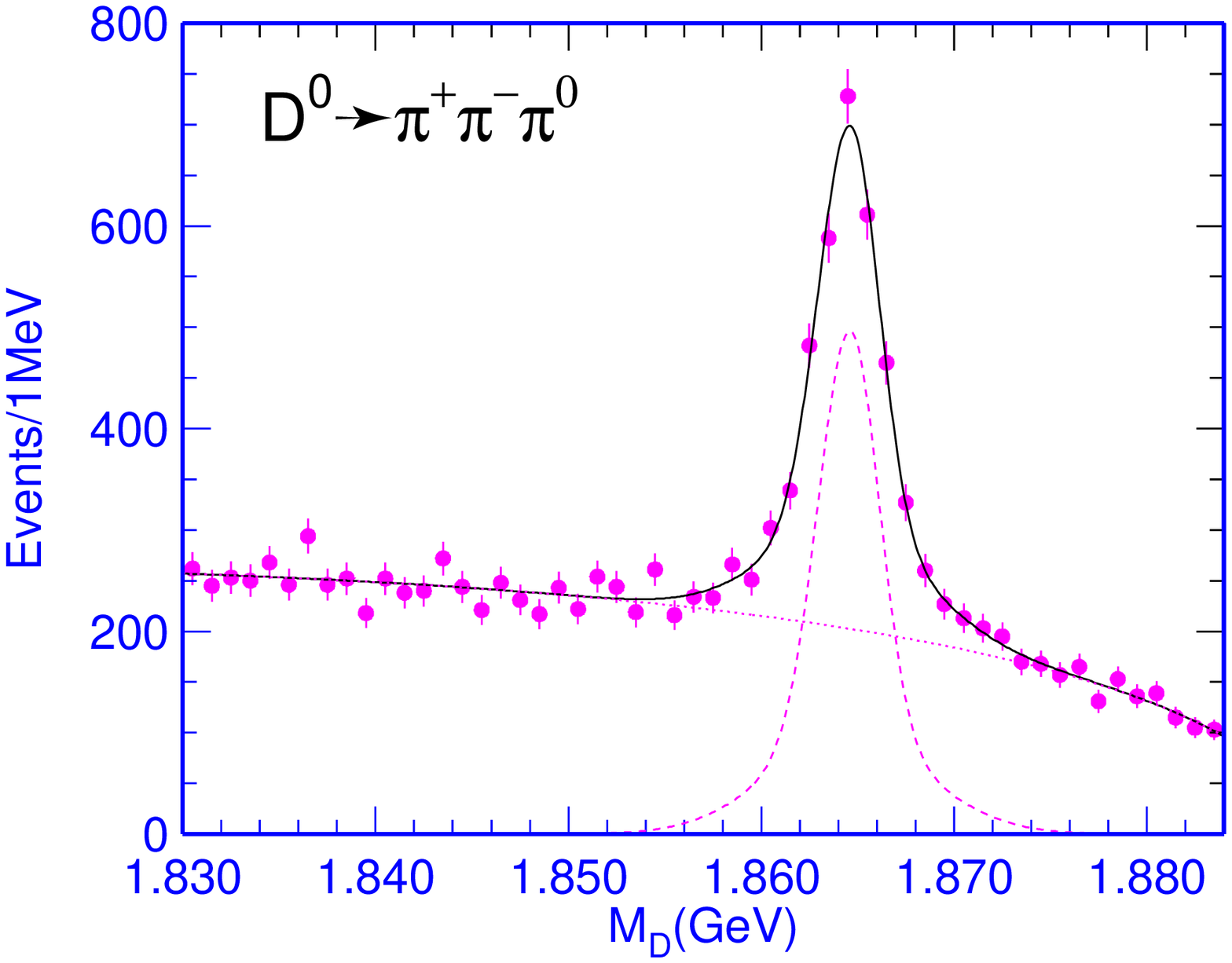}
\epsfxsize=0.33\textwidth\epsffile{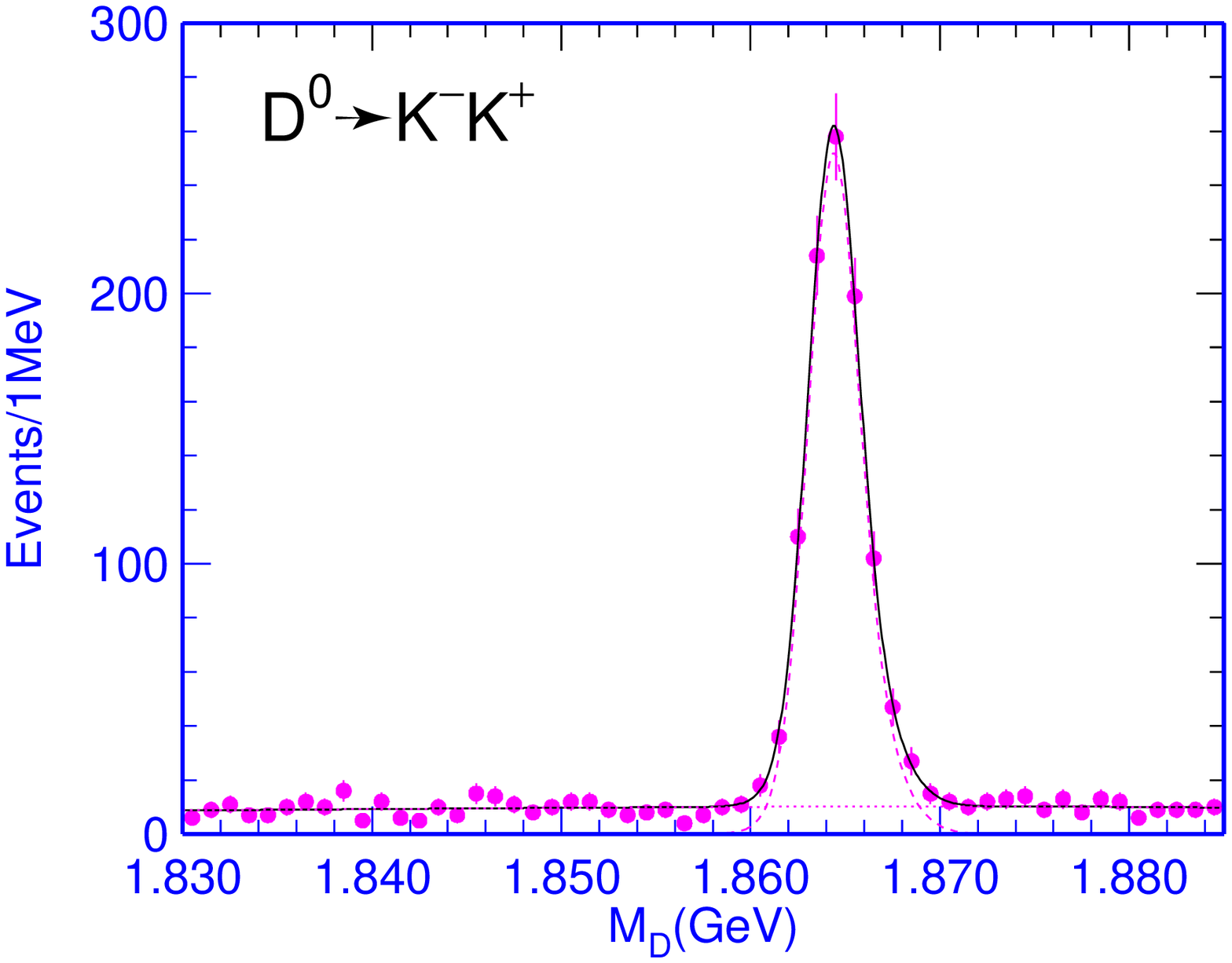}}}

\caption{Fits to the beam-constrained masses for different 
         fully reconstructed $D^{0}$ decay modes.
         The signal is described by a Gaussian and a bifurcated 
         Gaussian to account for the initial state radiation.
         The background is described by an Argus 
         function~\cite{argus}.}  
\label{md0bc1} 
\end{figure}

Full details of the hadronic event reconstruction and tag selection 
can be found in a separate paper~\cite{cleoc-Dtagging}.  
The selection is based on two variables.  
$\Delta E$ is the difference between the beam energy and the 
energy of the fully reconstructed $D^{0}$ candidate.  
$M_{D}$ is the beam-constrained mass of the $D^{0}$ candidate, 
which is defined as $M_{D} \equiv \sqrt{E_b^2 - | p_{D} |^2}$, 
where $E_{b}$ is the beam energy and $p_{D}$ is the measured 
momentum of the $D^{0}$ candidate.
Fits to the beam-constrained mass distributions for $D^{0}$ 
candidates are shown in Fig.~\ref{md0bc1}.  
Multiple combinations have been eliminated by selecting the 
candidate with the minimum value of $|\Delta E|$.  
The signal component in these fits consists of a Gaussian 
and a bifurcated Gaussian, which is found to adequately 
describe initial-state radiation (ISR).  The background component
in these fits  is presented by an Argus function~\cite{argus}.  
The tag yields and efficiencies for selecting tags are given in 
Table~\ref{table}.

\begin{table}
\caption{The tag yields, efficiencies, and signal yields and efficiencies
for $\dke$.} 
\begin{center}
\begin{tabular}{lccccc}\hline
Tags: $D^0\to$    &  tag yields  & $\epsilon_{\rm tag}$ 
  & signal yields & $\epsilon_{\rm signal}$ & ${\cal B}(\%)$ \\ \hline 
$\kpi$     & 10183$\pm$112    & (65.35$\pm$0.12)\%  
           & 245.7$\pm$15.3   & (42.69$\pm$0.21)\% & 3.69$\pm$0.23\\ 
$\kpp$     & 17208$\pm$160    & (33.53$\pm$0.07)\% 
           & 407.7$\pm$20.4   & (22.07$\pm$0.18)\% & 3.60$\pm$0.18 \\
$\kppp$    &  4475$\pm$208    & (16.10$\pm$0.06)\%  
           & 106.6$\pm$10.6   & (10.86$\pm$0.13)\% & 3.53$\pm$0.39 \\
$\kpppc$   & 14515$\pm$162    & (46.93$\pm$0.09)\% 
           & 310.3$\pm$18.0   & (29.80$\pm$0.20)\% & 3.37$\pm$0.20\\ 
$\kspp$    &  4629$\pm$101    & (37.05$\pm$0.15)\%  
           & 100.0$\pm$10.2   & (24.25$\pm$0.19)\% & 3.30$\pm$0.34 \\ 
$\ksppp$   &  4403$\pm$240    & (18.51$\pm$0.09)\%  
           & 119.2$\pm$10.6   & (12.83$\pm$0.19)\% & 3.91$\pm$0.41 \\ 
$\ksp$     &  1502$\pm$43     & (27.92$\pm$0.22)\%   
           &  26.6$\pm$5.3    & (19.19$\pm$0.17)\% & 2.58$\pm$0.52 \\ 
$\ppp$     &  2533$\pm$176    & (39.91$\pm$0.19)\%   
           &  63.3$\pm$8.0    & (26.67$\pm$0.19)\% & 3.74$\pm$0.54 \\ 
$\kk$      &   928$\pm$35     & (56.08$\pm$0.38)\%
           &  22.4$\pm$4.9    & (37.04$\pm$0.21)\% & 3.66$\pm$0.81 \\ 
\hline
all tags & & & 1405.1$\pm$38.5 & weighted ave. & 3.52$\pm$0.10 \\ 
\hline
\end{tabular} 
\end{center}
\label{table}
\end{table}

The fully reconstructed $D^{0}$ meson serves as a tag and 
results in great suppression of backgrounds. 
We then identify an electron and a set of hadrons recoiling against 
the tag, to reconstruct the semileptonic decay side of the
$D^{0}\bar {D^{0}}$ system.
Electron candidates are required to have $| \cos{\theta} |$ $<$ 0.9, 
where $\theta$ is the angle between the electron direction and the 
beam axis, with momenta greater than 200 MeV, the minimum to reach
the CsI calorimeter for energy measurement.
Electron candidates are selected with criteria that rely mostly
on the ratio of the energy deposited in the CsI calorimeter to the
measured momentum ($E/p$), $dE/dx$ and RICH information.
Extensive studies of efficiencies and misidentification rates
have been performed using CLEO-c data for this electron 
identification procedure.
Charged pions and kaons are identified using $dE/dx$ and RICH
information. 
For charged tracks with momenta less than 0.6 GeV,
consistent $dE/dx$ information (within 3 standard deviation 
($\sigma$)) is required for pion or kaon candidates.
For charged tracks with momenta greater than 0.6 GeV,
additional requirement on RICH, if available, is imposed for
pion or kaon candidates.
$K^{-}$ and $\pi^{0}$ candidates are combined to form $K^{*-}$
candidates in  $D^0 \to \kste$ decays. We require the invariant
masses of the $K^{*-}$ candidates to be within 100 MeV of the
$K^{*-}$ nominal mass~\cite{PDG}. In case of multiple $K^{*-}$ 
candidates, we select the candidate whose mass is closest to the 
$K^{*-}$ nominal mass.
$\pi^{-}$ and $\pi^{0}$ candidates are combined to form $\rho^{-}$
candidates in  $D^0 \to \rhoe$ decays. We require the invariant
masses of the $\rho^{-}$ candidates to be within 150 MeV of the
$\rho^{-}$ nominal mass~\cite{PDG}. In case of multiple $\rho^{-}$ 
candidates, we select the candidate whose mass is closest to the 
$\rho^{-}$ nominal mass.
We reconstruct exclusive $D^0$ semileptonic decays 
into $\ke$, $\pie$, $\kste$ and $\rhoe$. 
No additional charged track is allowed to exist, besides those
of the tagged $D^{0}$ and the semileptonic side of the other 
$\bar {D^{0}}$.
The excellent particle identification of the CLEO-c detector
helps in this reconstruction. The unique kinematics of 
threshold production provide additional and very powerful means to 
reject background from misidentified and missing particles. The 
difference between the missing energy and missing momentum in a 
event, $U=E_{\rm miss}-p_{\rm miss}$, would peak at zero if the event is 
correctly reconstructed due to the undetected neutrino.

\begin{figure}[htbp]
\epsfxsize=0.50\textwidth\centerline{\mbox{\epsffile{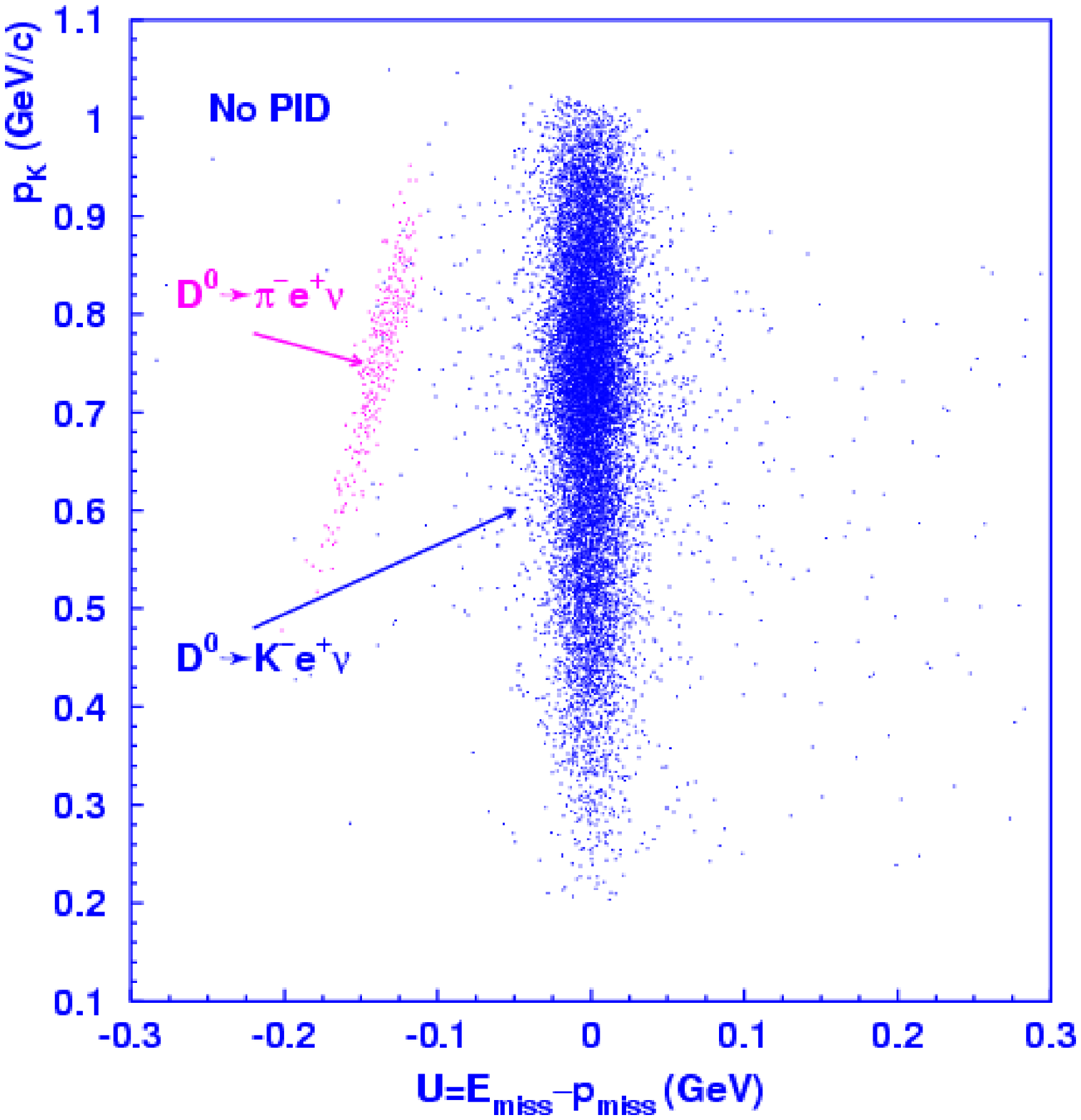}
\epsfxsize=0.50\textwidth\mbox{\epsffile{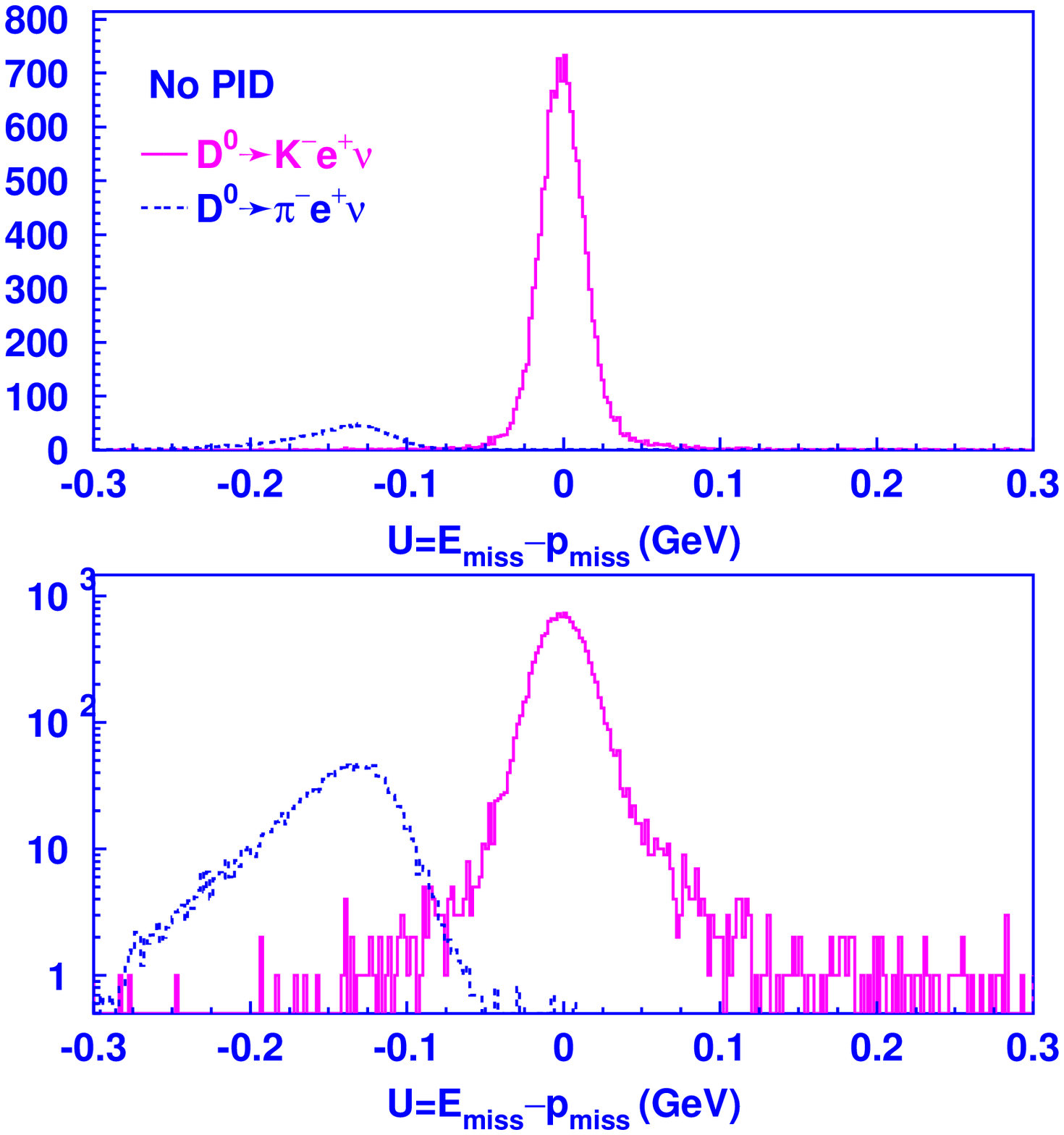}}}}
\caption{Kaon momentum vs. $U=E_{\rm miss}-p_{\rm miss}$ from
         $\dke$ signal MC. The background from $\dpie$ reconstructed 
         as $\dke$ is overlaid and normalized by 
         ${{\cal B}(\dpie)\over {\cal B}(\dke)}=0.1$ from PDG~\cite{PDG}. 
         The right plot shows the projection onto the $U$ axis in 
         linear and log scales.}
\label{pku2d}
\end{figure}

In Fig.~\ref{pku2d},
we present kaon momentum versus $U$ for signal $\dke$ and the 
background from $\dpie$ reconstructed as $\dke$
without hadron identification using MC simulation. It clear 
that the $\dke$ signal and $\dpie$ background are well-separated
even without kaon identification, thanks to the kinematics of threshold
production. The separation decreases as the kaon momentum increases.

\begin{figure}[htbp]
\epsfxsize=0.50\textwidth\centerline{\mbox{\epsffile{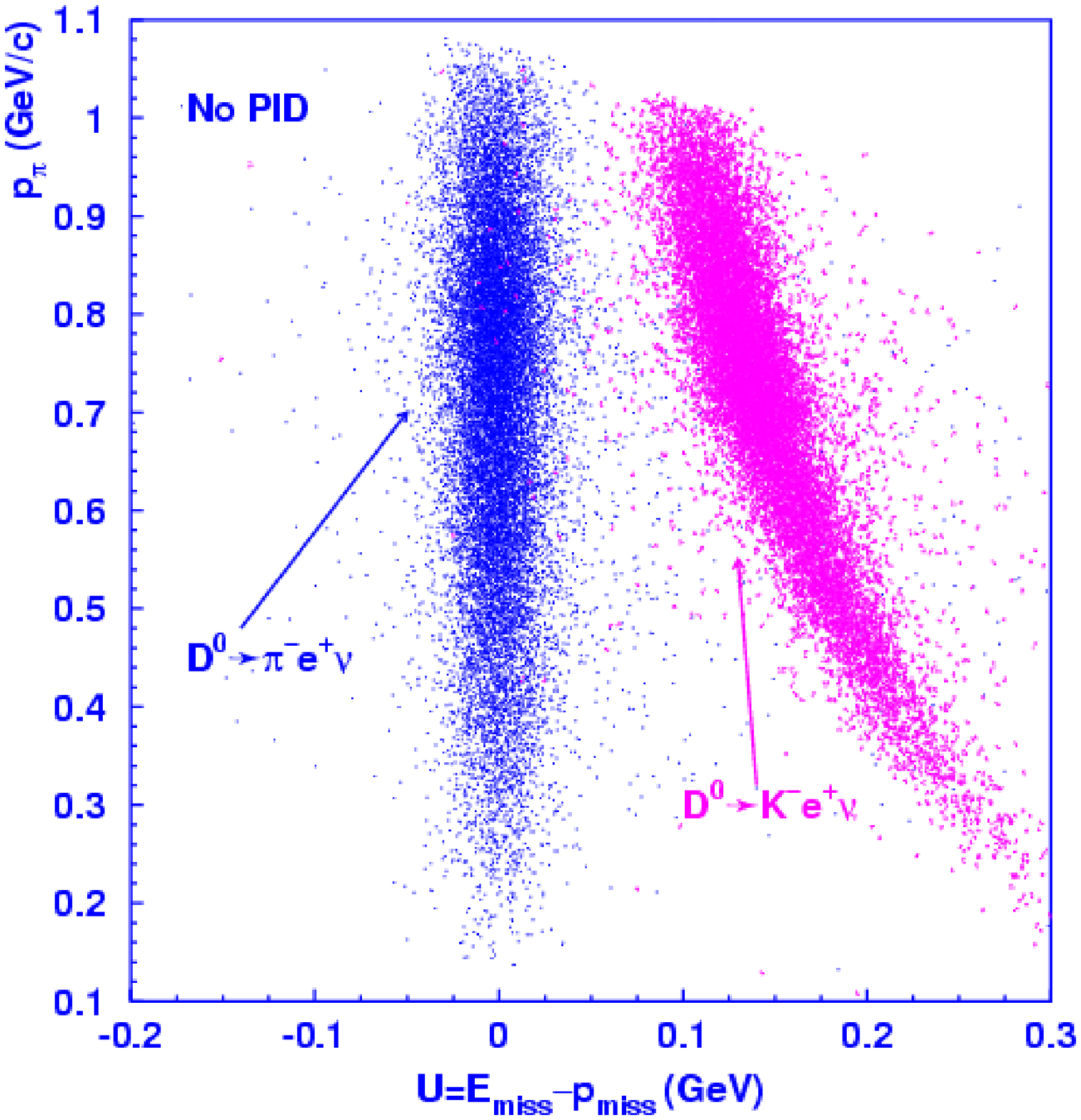}
\epsfxsize=0.50\textwidth\mbox{\epsffile{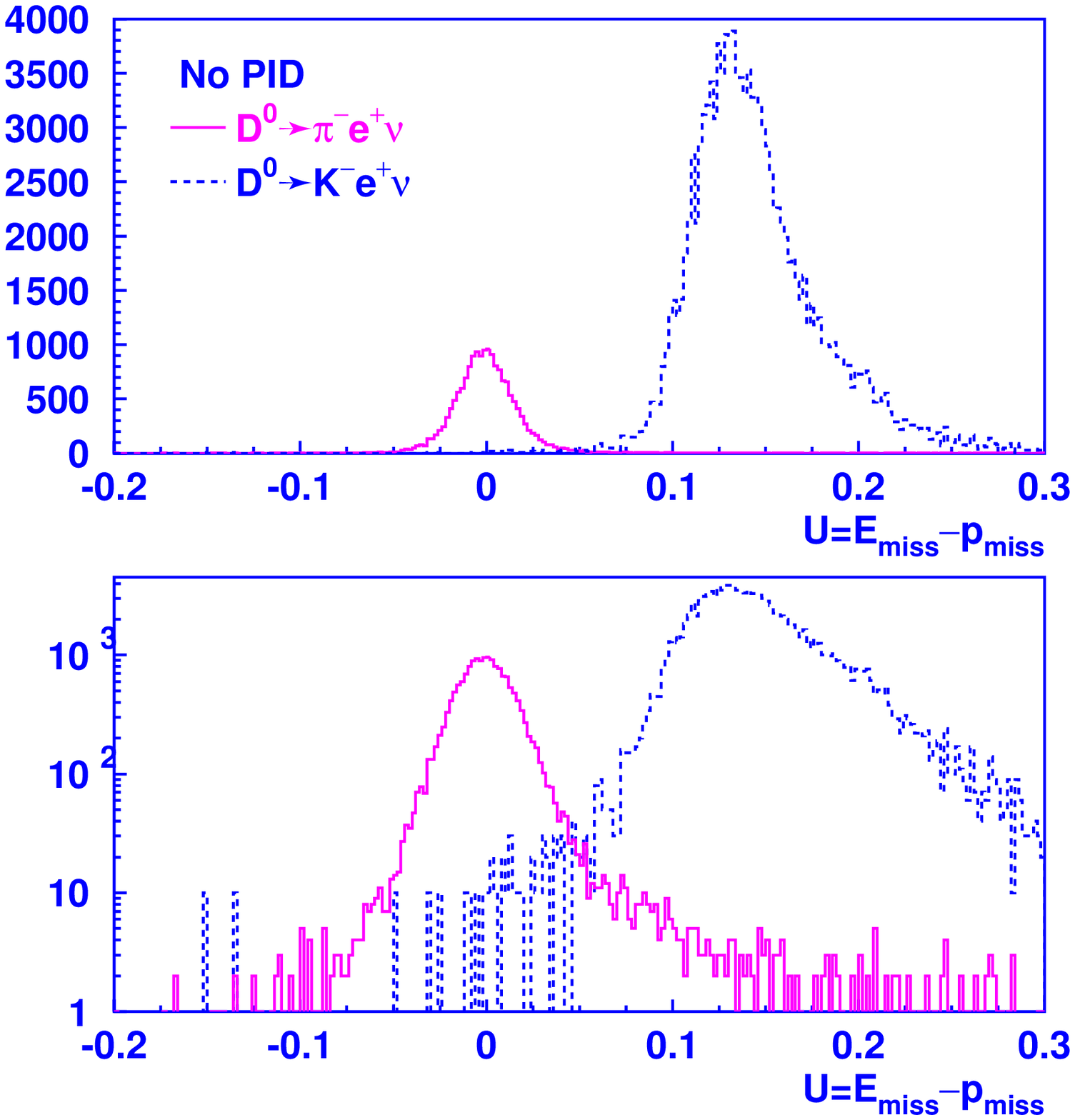}}}}
\caption{Pion momentum vs. $U=E_{miss}-p_{miss}$ from 
         $\dpie$ signal MC. The background from $\dke$ reconstructed 
         as $\dpie$ is overlaid and normalized by 
         ${{\cal B}(\dpie)\over {\cal B}(\dke)}=0.1$ from PDG~\cite{PDG}. 
         The right plot shows the projection onto the $U$ axis in 
         linear and log scales.}
\label{ppiu2d}
\end{figure}

Similarly, in Fig.~\ref{ppiu2d}, we present pion momentum versus 
$U$ for signal $\dpie$ and the background from $\dke$ reconstructed 
as $\dpie$ without hadron identification. Again, the signal and the 
dominant background are well-separated. 
Therefore, tight hadron identification is not necessary for the  
semileptonic decay (signal) side. 

\section{RESULTS} 

The absolute branching fractions of $D^{0}$ semileptonic decay is
given by
\begin{eqnarray}
{\cal B}={N_{\rm signal}/\epsilon_{\rm signal}\over 
          N_{\rm tag}/\epsilon_{\rm tag}}.  
\label{ourbr} 
\end{eqnarray} 
$N_{\rm signal}$ is the number of the $D^{0} \bar{D^{0}}$ events
with one $D^{0}$ fully reconstructed and the other $\bar{D^{0}}$
reconstructed through its corresponding exclusive semileptonic
decay mode. $\epsilon_{signal}$ is the efficiency for 
constructing the fully reconstructed (tag) $D^{0}$ and the 
exclusive semileptonic decay of the $\bar {D^{0}}$.
Similarly,  $N_{\rm tag}$ is the number of the fully reconstructed
(tag) $D^{0}$ events observed, and $\epsilon_{tag}$ is the
efficiency for fully reconstructing the $D^{0}$.
Therefore, $\epsilon_{signal}$/$\epsilon_{tag}$ is the 
efficiency for reconstructing the exclusive $D^{0}$ 
semileptonic decay in the presence of a found $\bar {D^{0}}$ tag.

\begin{figure}[htbp]
\centerline{\mbox{\epsfxsize=0.50\textwidth\epsffile{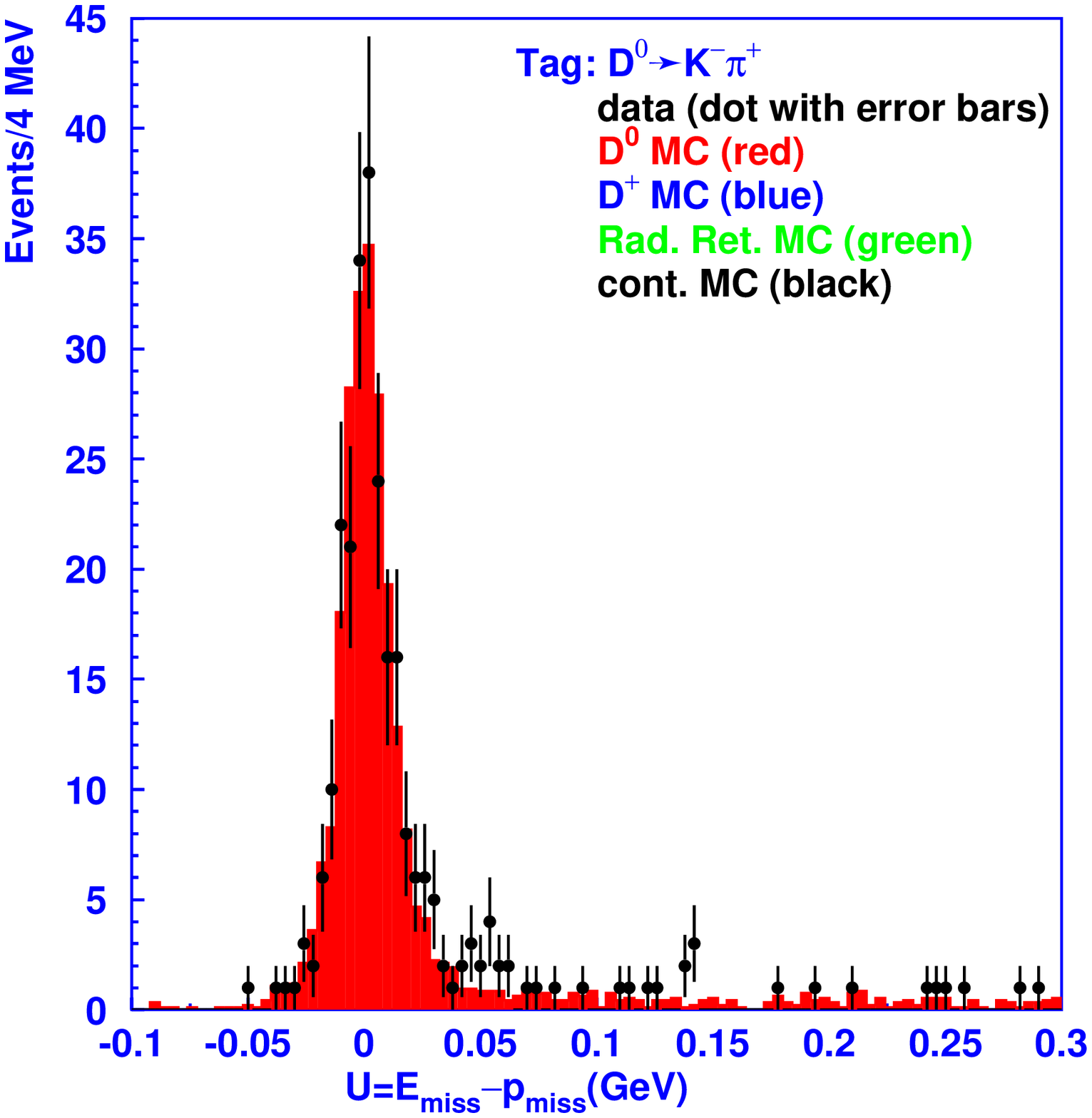}
\epsfxsize=0.50\textwidth\epsffile{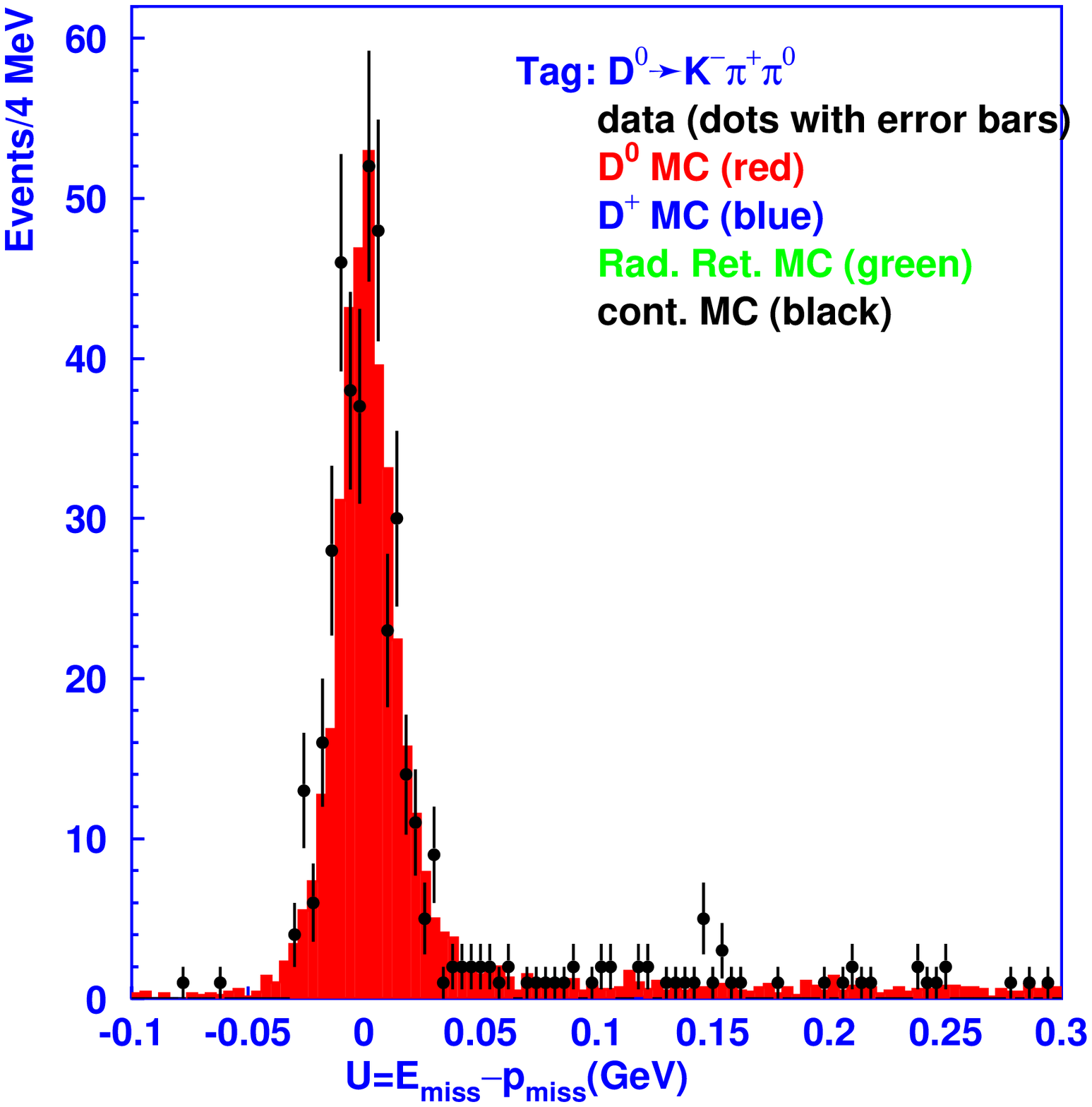}}} 
\centerline{\mbox{\epsfxsize=0.50\textwidth\epsffile{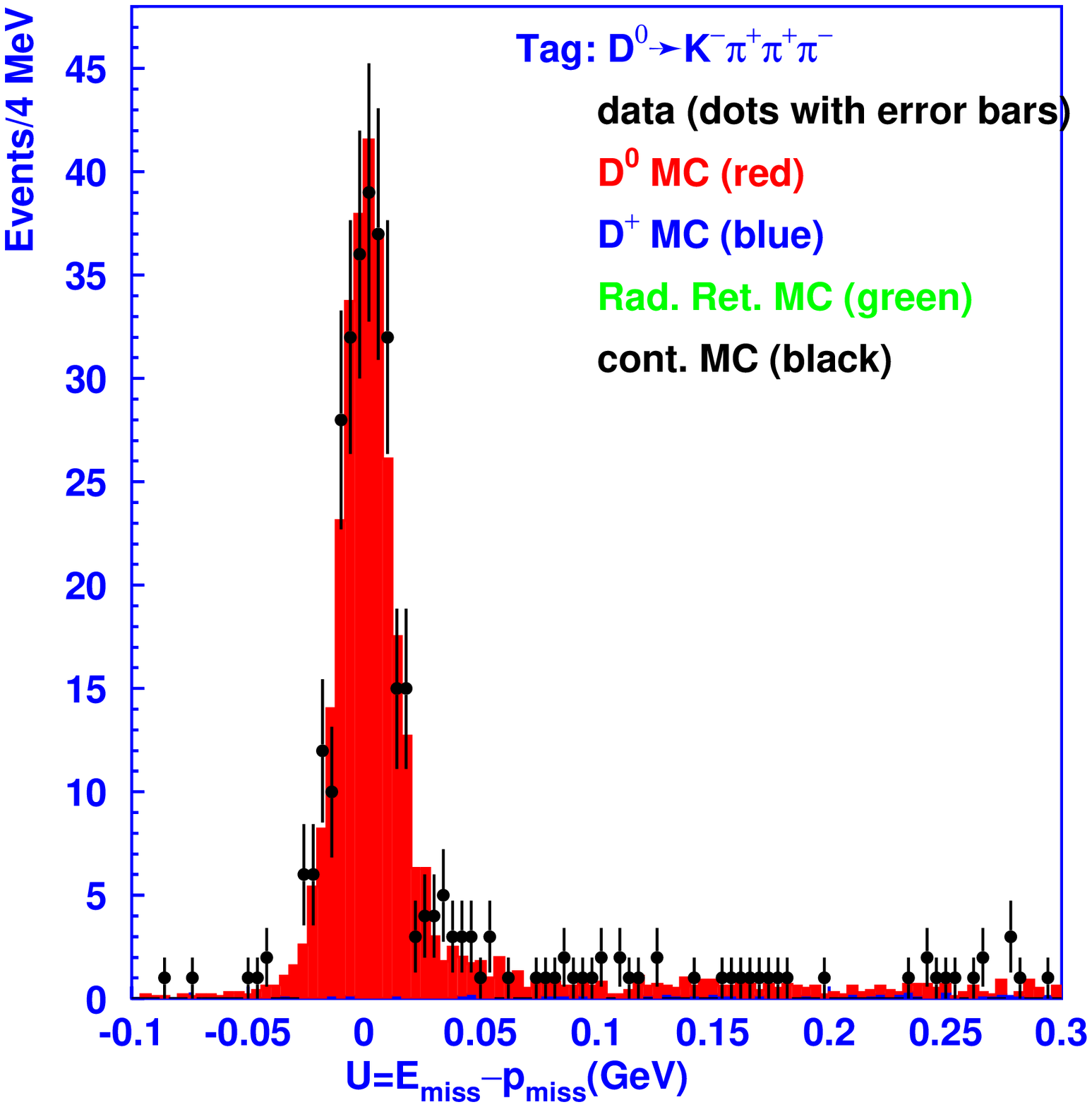}
\epsfxsize=0.50\textwidth\epsffile{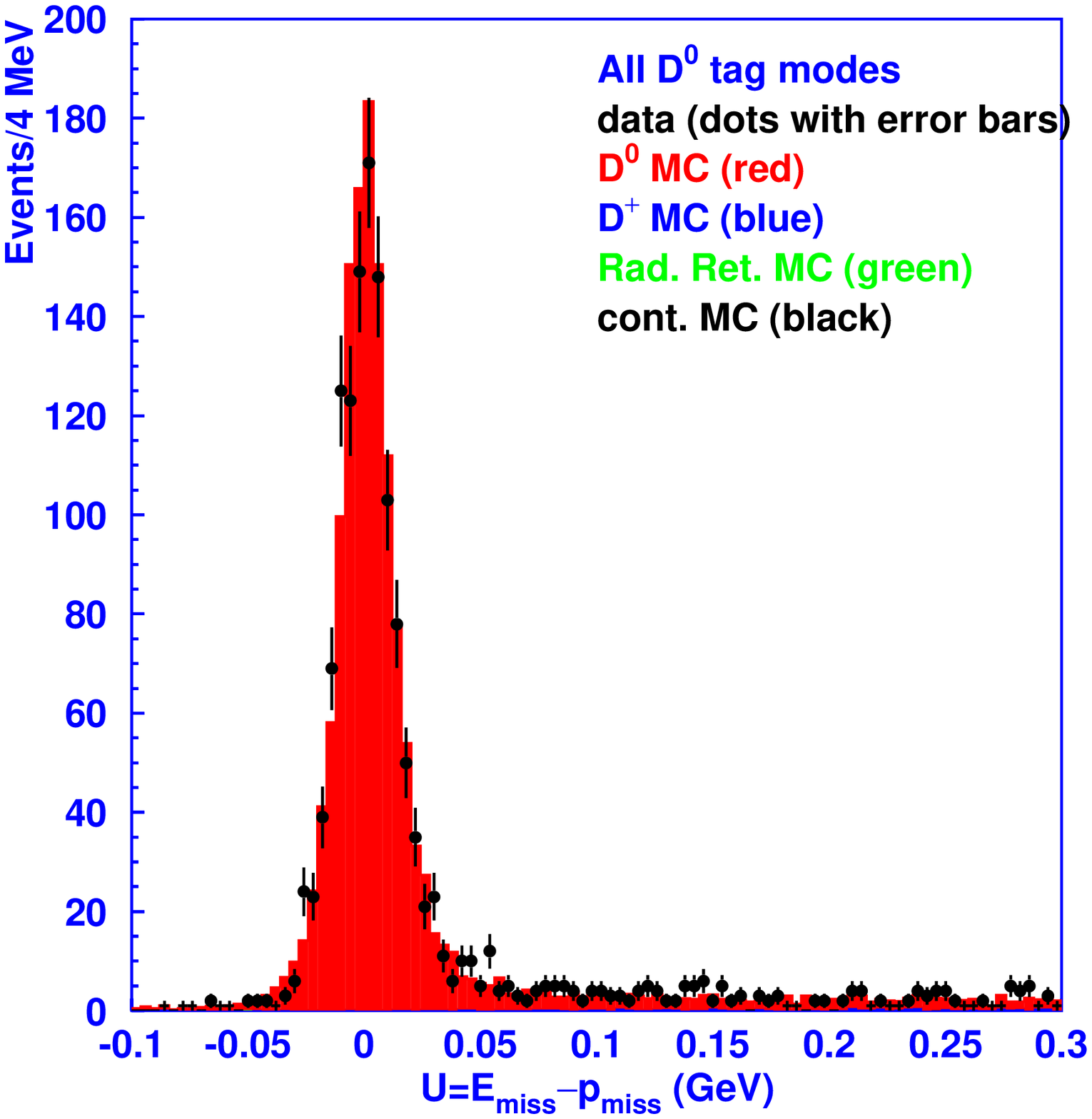}}} 
\caption{Comparison of $U=E_{\rm miss}-p_{\rm miss}$ for $\dke$ 
         between the data and MC.}
\label{uklnu}
\end{figure}

\begin{figure}[htbp]
\centerline{\mbox{\epsfxsize=0.50\textwidth\epsffile{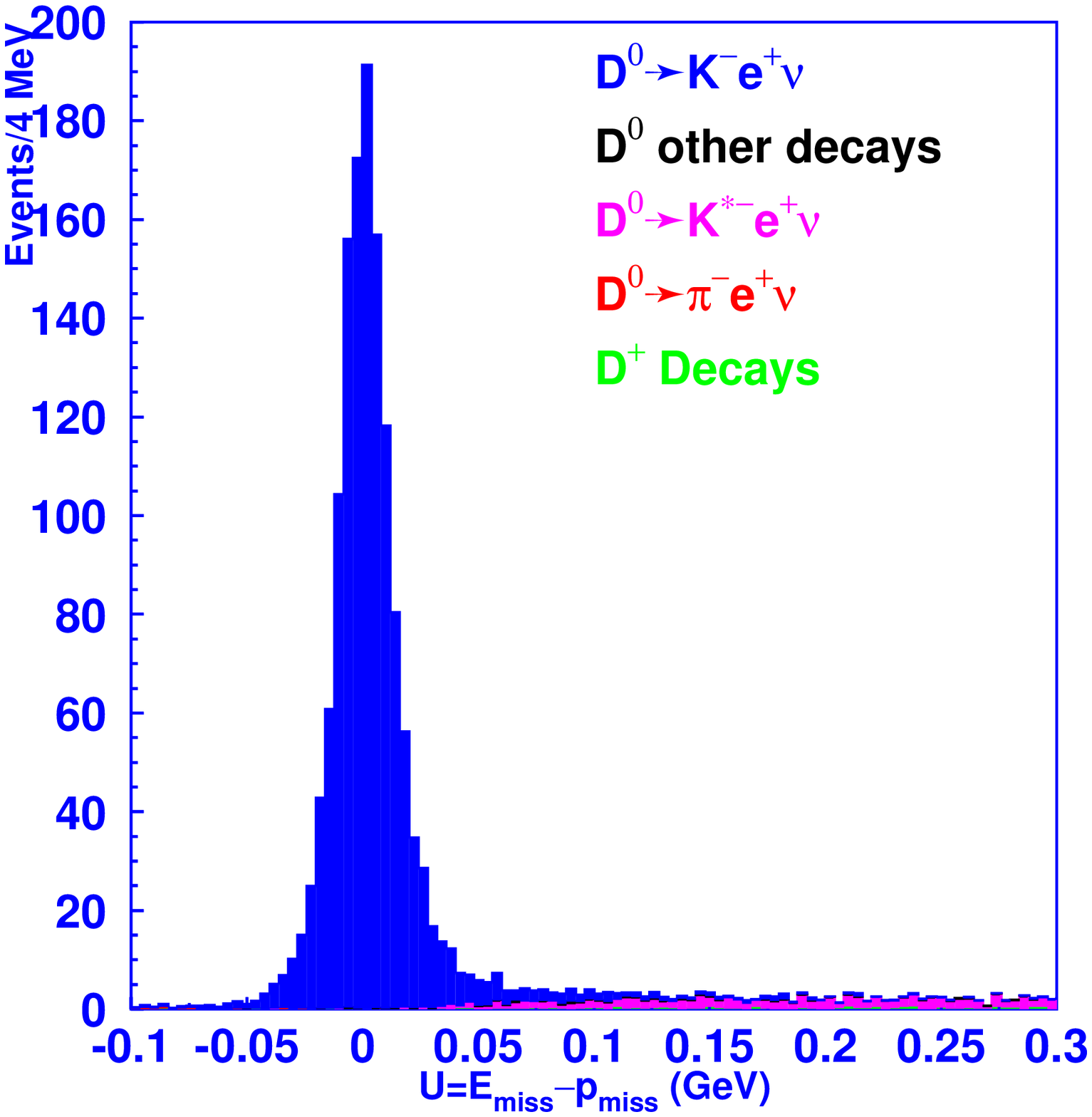}
\epsfxsize=0.50\textwidth\epsffile{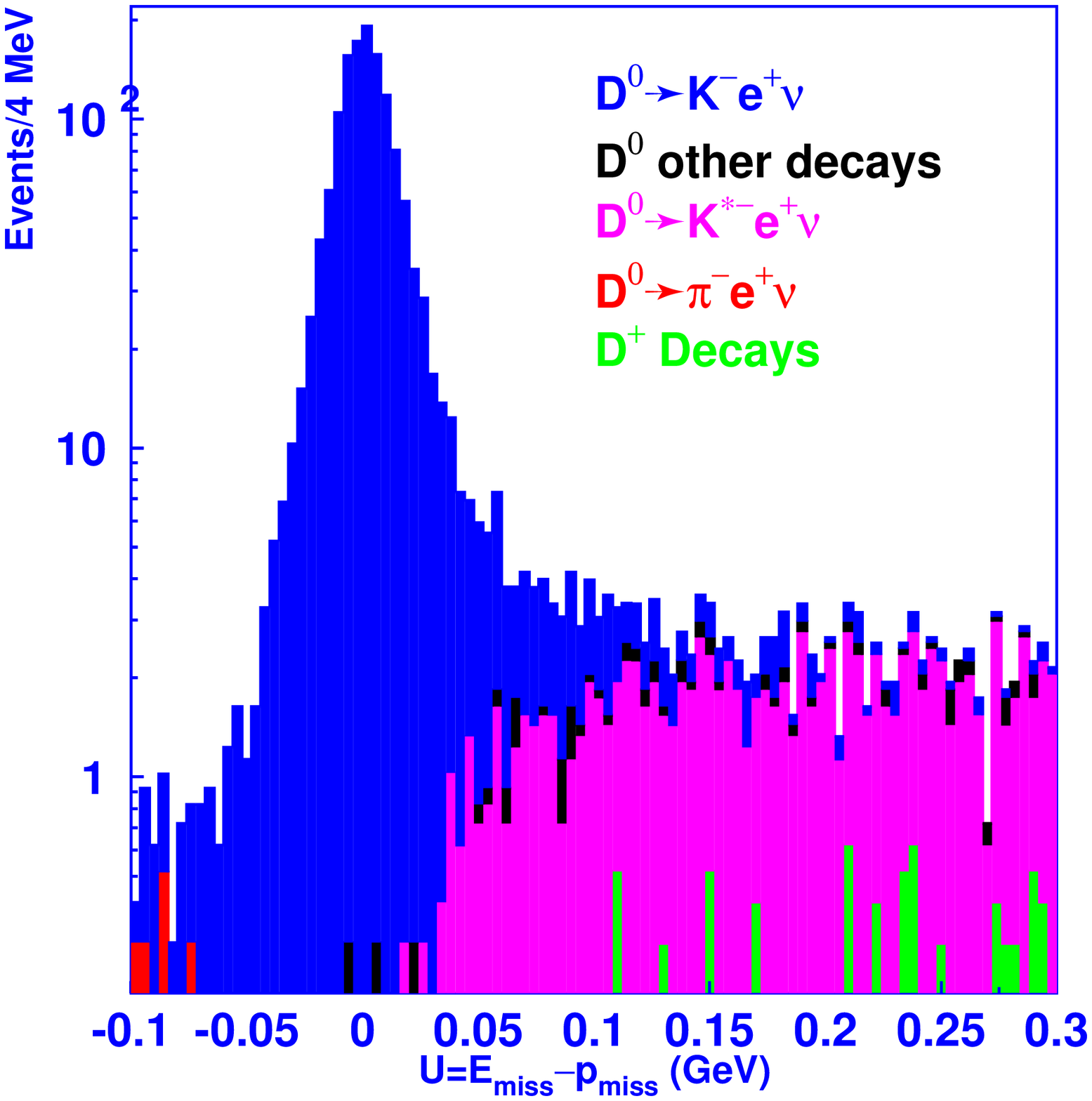}}} 
\caption{Decomposition of $U=E_{\rm miss}-p_{\rm miss}$ for $\dke$ 
         from MC for all tag modes combined.}
\label{uklnu-dec}
\end{figure}

In Fig.~\ref{uklnu}, we present the $U$ distributions from data
and MC for the selected $\dke$ events with different tag modes. 
The comparison shows good agreement between the data and MC. 
Fig.~\ref{uklnu-dec} shows the decomposition of the $U$ 
distributions from MC for all tag modes combined.
The fit to the data is shown in Fig.~\ref{signal-yield}.
The $D^0 \to \ke$ yields from various $D^{0}$ tag modes are given in 
Table~\ref{table}. The efficiencies and branching fractions are also 
listed in Table~\ref{table} for different tag modes.

\begin{figure}[htbp]
\centerline{\mbox{\epsfxsize=0.48\textwidth\epsffile{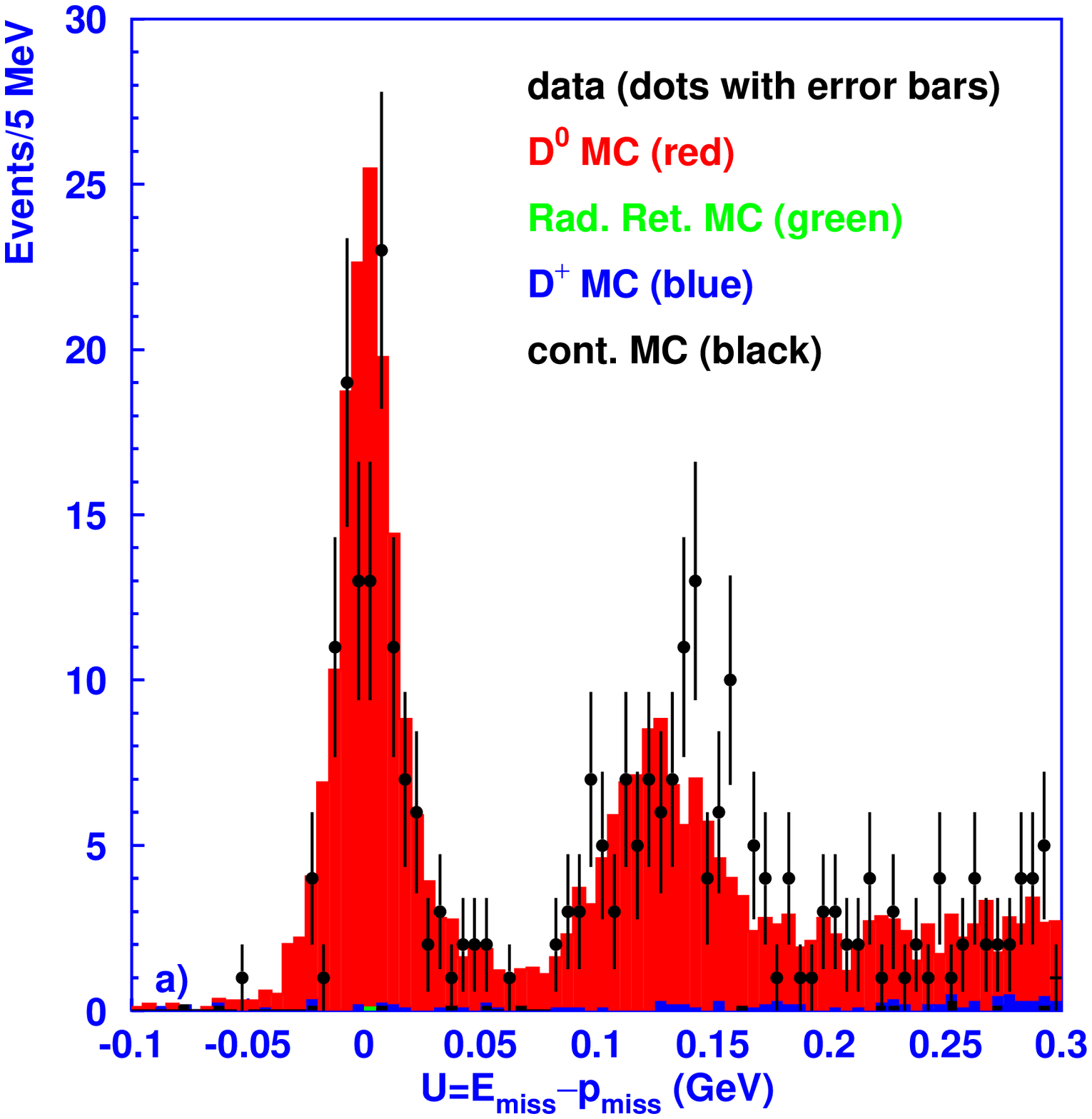}
\epsfxsize=0.48\textwidth\epsffile{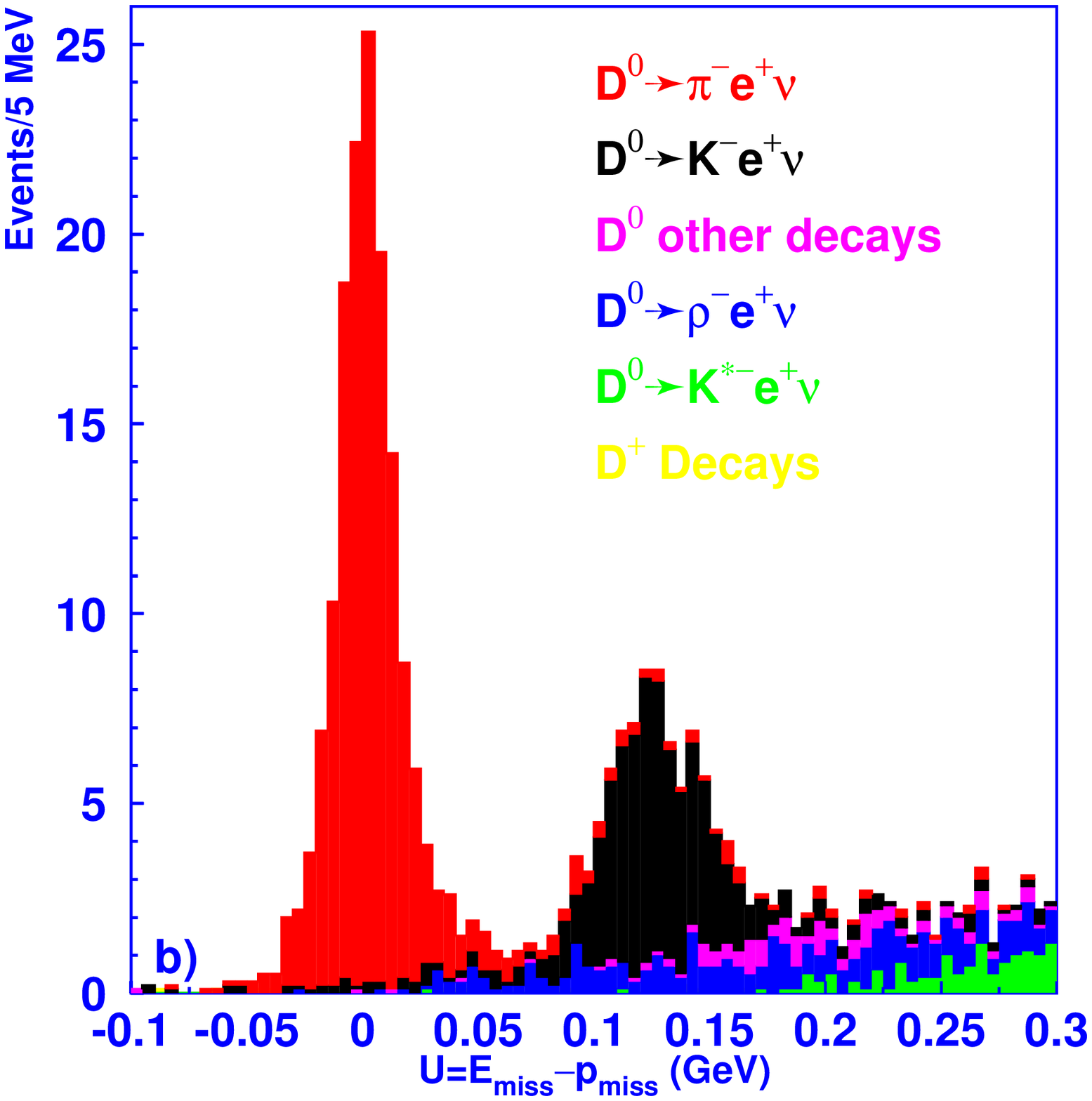}}} 
\caption{$U=E_{\rm miss}-p_{\rm miss}$ for the selected $\dpie$ 
         events (left), and the background components (right). } 
\label{upil}
\end{figure}

In Fig.~\ref{upil}, we present the $U$ distribution from data and MC 
for the selected $\dpie$ events with all tag modes combined due to 
limited statistics. 
The comparison shows reasonable agreement between
data and MC. There is a clear excess in the $\dpie$ signal region 
(near zero) in data which is in good agreement with MC prediction. 
The background decomposition is also shown in Fig.~\ref{upil}. 
We fit the $U$ distribution with a Gaussian, which corresponds 
to the $\dpie$ signal near zero, another Gaussian which corresponds to
the $\dke$ background on the right side, and a 2nd order polynomial
function. The 2nd order polynomial is determined from MC simulation.  
The fit to the data is shown in Fig.~\ref{signal-yield}.
We observe 109.1$\pm$10.9 events. The signal efficiencies and 
branching fractions are given in Table~\ref{table2}.  

\begin{table}[hbtp] 
\caption{Efficiencies and yields for $\dpie$, $\dkste$ and 
$\drhoe$ using different tag modes.} 
\begin{center}
\begin{tabular}{lccc}\hline 
Signal Modes &  $\dpie$    &    $\dkste$   &  $\drhoe$ \\ 
Tags       & $\epsilon$   & $\epsilon$ & $\epsilon$ \\ \hline 
$\dkpi$    & (48.06$\pm$0.30)\% 
           & (14.34$\pm$0.21)\% 
           & (18.13$\pm$0.22)\% \\
$\dkpp$    & (24.93$\pm$0.26)\% 
           &  (6.97$\pm$0.15)\% 
           &  (8.90$\pm$0.16)\% \\ 
$\dkppp$   & (12.47$\pm$0.20)\% 
           &  (3.51$\pm$0.11)\% 
           &  (4.52$\pm$0.12)\% \\
$\dkpppc$  & (33.28$\pm$0.29)\% 
           &  (9.61$\pm$0.17)\% 
           & (11.62$\pm$0.19)\% \\
$\dkspp$   & (25.77$\pm$0.27)\% 
           &  (7.99$\pm$0.16)\% 
           &  (9.45$\pm$0.17)\% \\
$\dksppp$  & (14.18$\pm$0.15)\% 
           &  (3.95$\pm$0.11)\% 
           &  (4.94$\pm$0.13)\% \\
$\dksp$    & (21.45$\pm$0.25)\% 
           &  (6.28$\pm$0.14)\% 
           &  (7.39$\pm$0.15)\% \\
$\dppp$    & (28.98$\pm$0.27)\% 
           &  (8.83$\pm$0.17)\% 
           & (10.64$\pm$0.18)\% \\ 
$\dkk$     & (40.14$\pm$0.29)\% 
           & (12.33$\pm$0.19)\% 
           & (15.14$\pm$0.21)\% \\ 
\hline
Yields     & 109.1$\pm$10.9 & 88.0$\pm$9.7 & $30.1\pm$5.8 \\ \hline 
${\cal B}(\times10^{-3})$  & (2.46$\pm$0.25) & 20.69$\pm$2.28  
           & $1.89\pm0.36$ \\ \hline 
\end{tabular} 
\end{center}
\label{table2}
\end{table}

\begin{figure}[htbp]
\centerline{\mbox{\epsfxsize=0.50\textwidth\epsffile{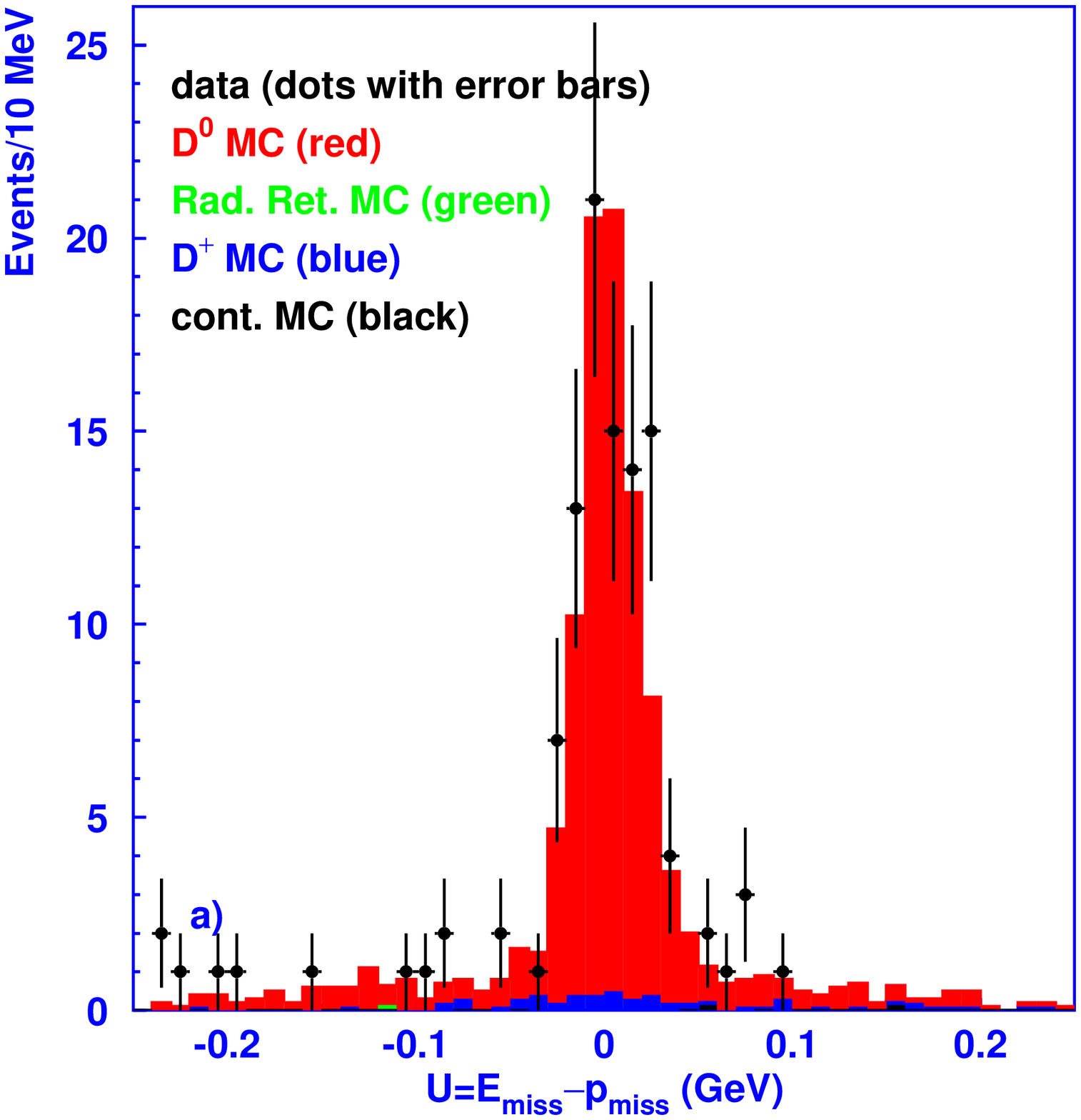} 
\epsfxsize=0.50\textwidth\epsffile{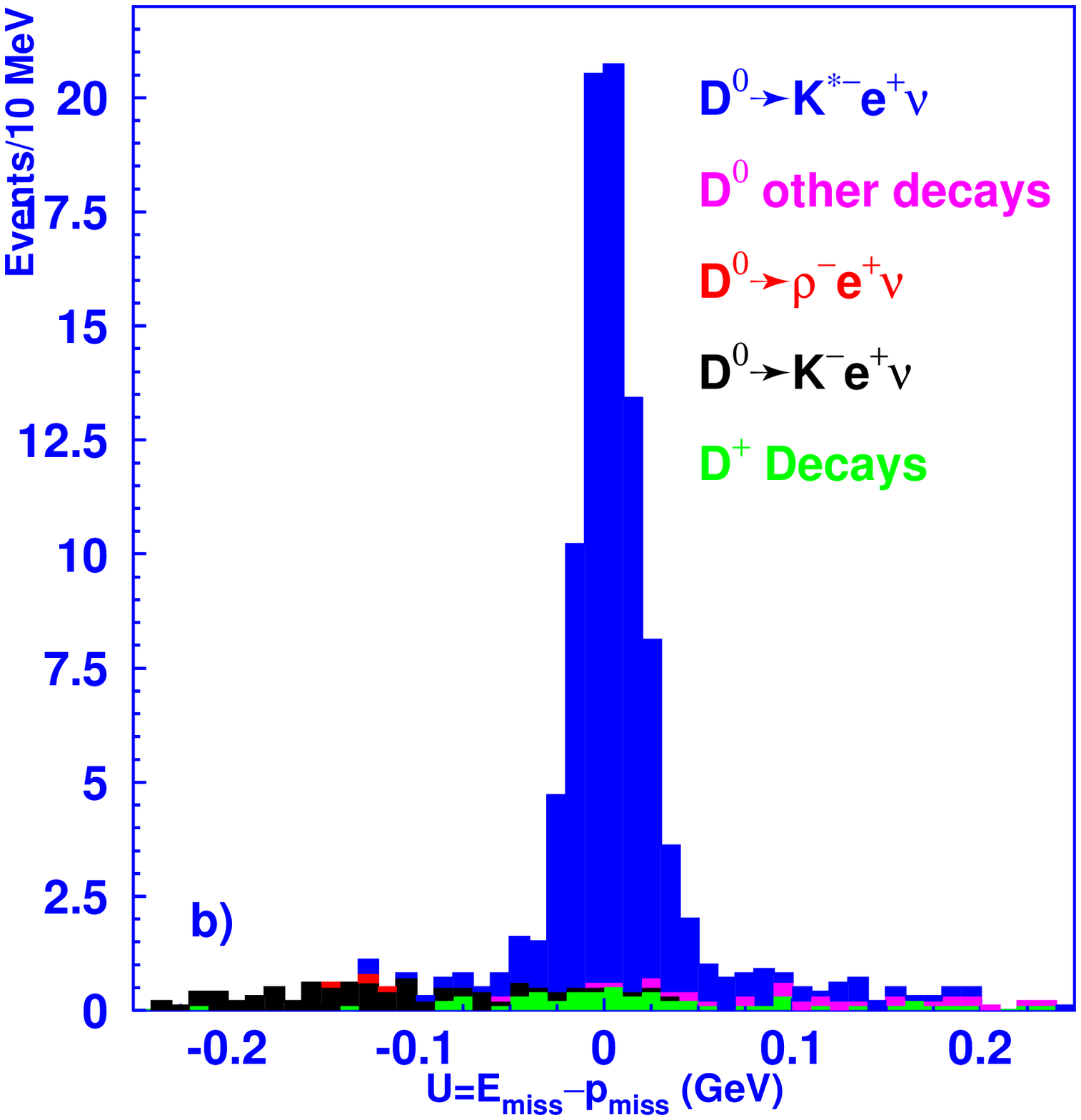}}} 
\caption{$U$ distribution for the selected $\dkste$ events (left) and 
         the components of backgrounds for $\dkste$ (right).} 
\label{ukst}
\end{figure}

In Fig.~\ref{ukst}, we present the comparison between the data and 
MC for $D^0 \to \kste$.
The comparison shows good agreement between data and MC.
The background decomposition is also shown in Fig.~\ref{ukst}. 
We fit the $U$ distribution with a Gaussian and a 2nd order 
polynomial determined from MC simulation. 
The fit to the data is shown in Fig.~\ref{signal-yield}.
The yields and efficiencies are given in Table~\ref{table2}.
  
\begin{figure}[htbp]
\centerline{\mbox{\epsfxsize=0.45\textwidth\epsffile{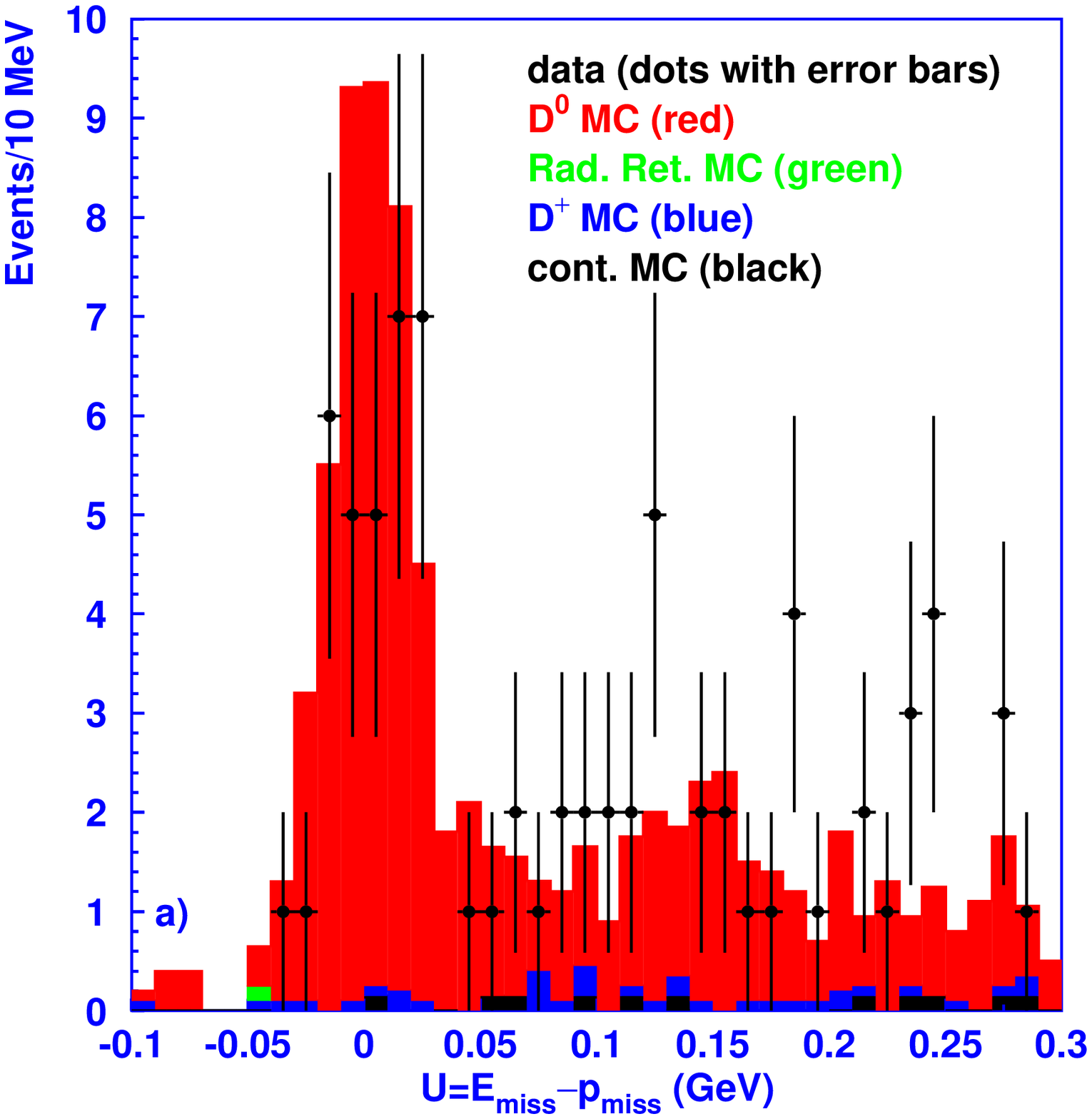}
\epsfxsize=0.45\textwidth\epsffile{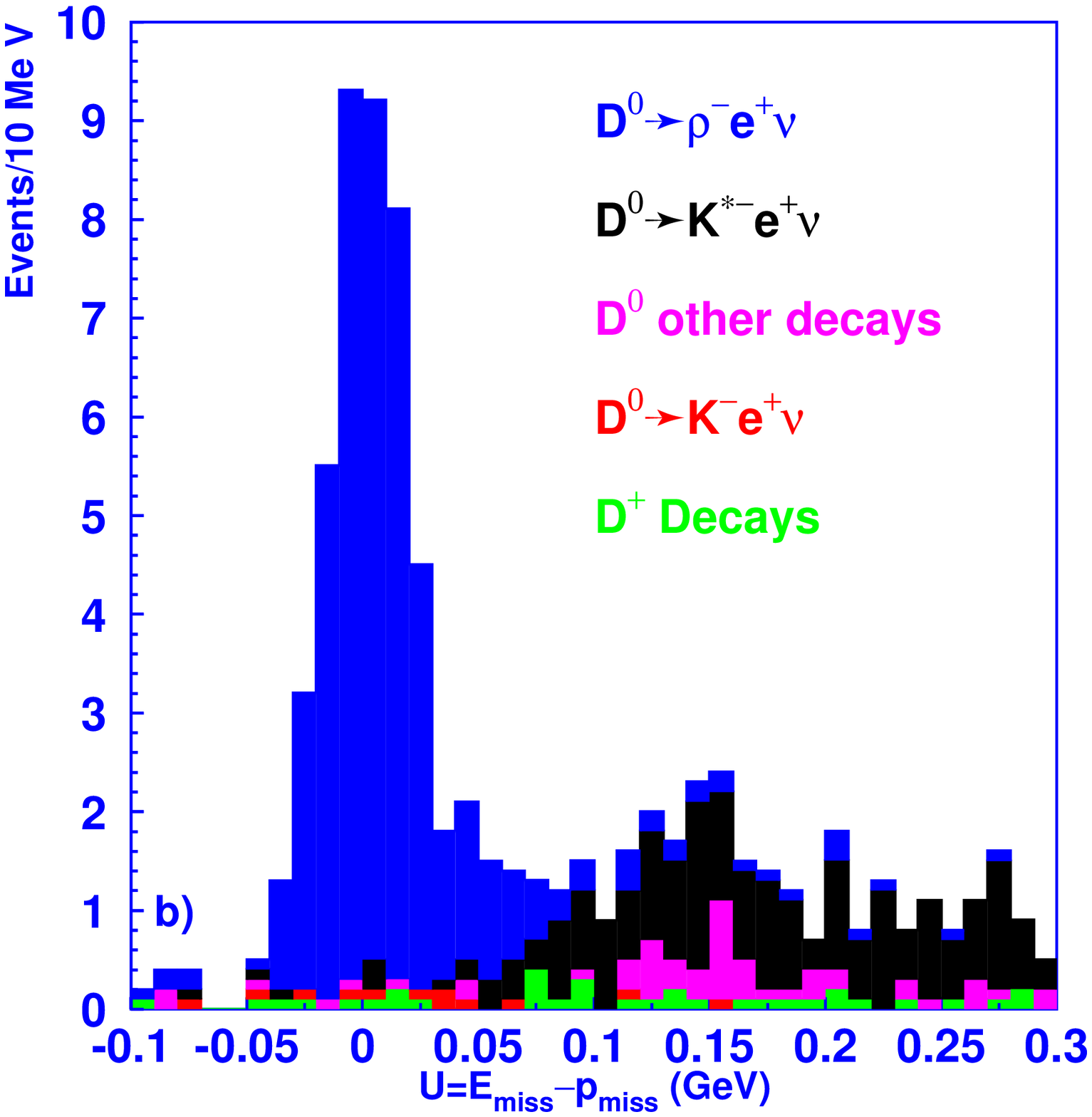}}} 
\caption{$U$ distribution for the selected  $\drhoe$ events (left) and 
the decomposition of backgrounds (right).} 
\label{urho}
\end{figure}

\begin{figure}[htbp]
\centerline{\mbox{\epsfxsize=0.50\textwidth\epsffile{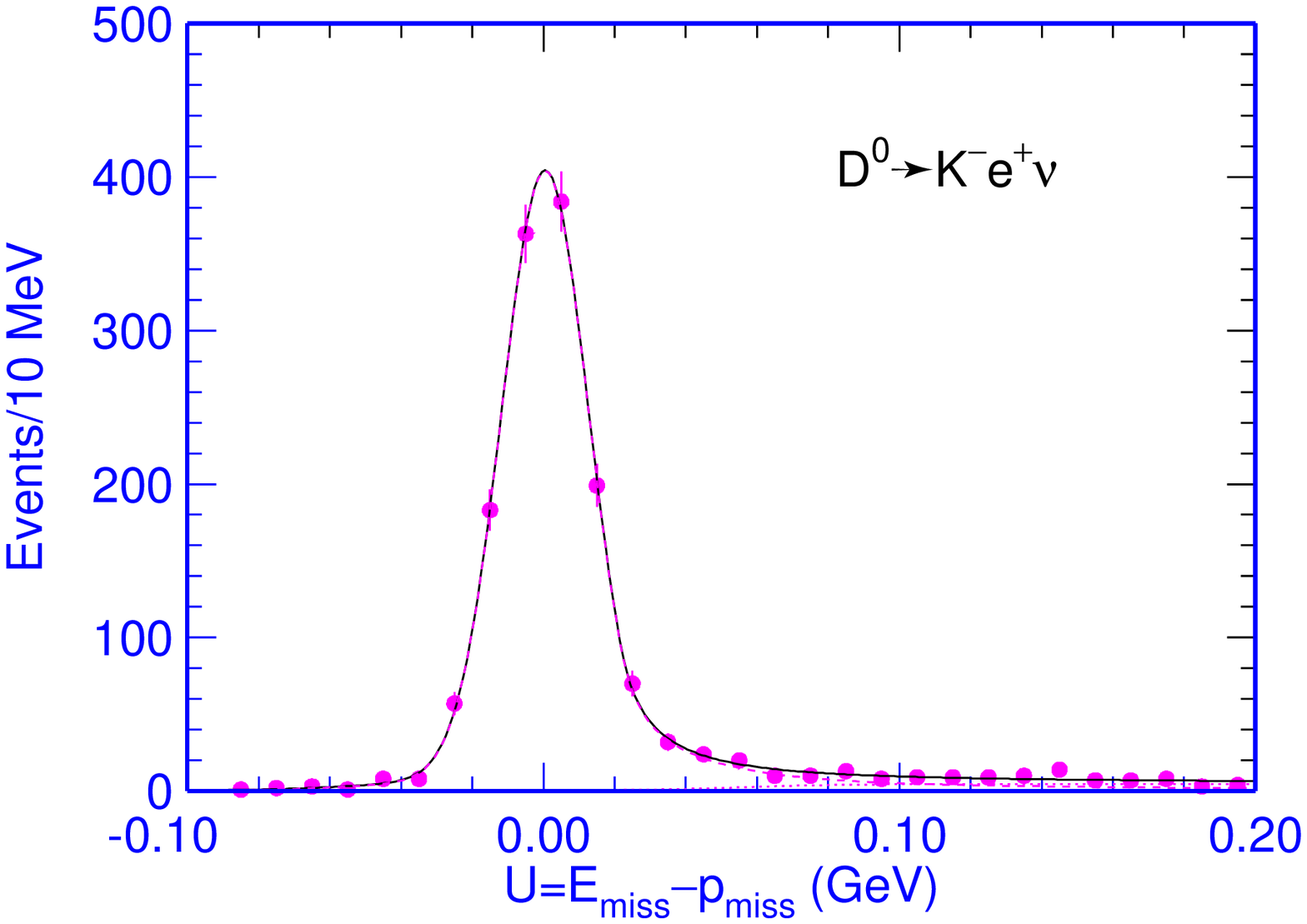}
\epsfxsize=0.50\textwidth\epsffile{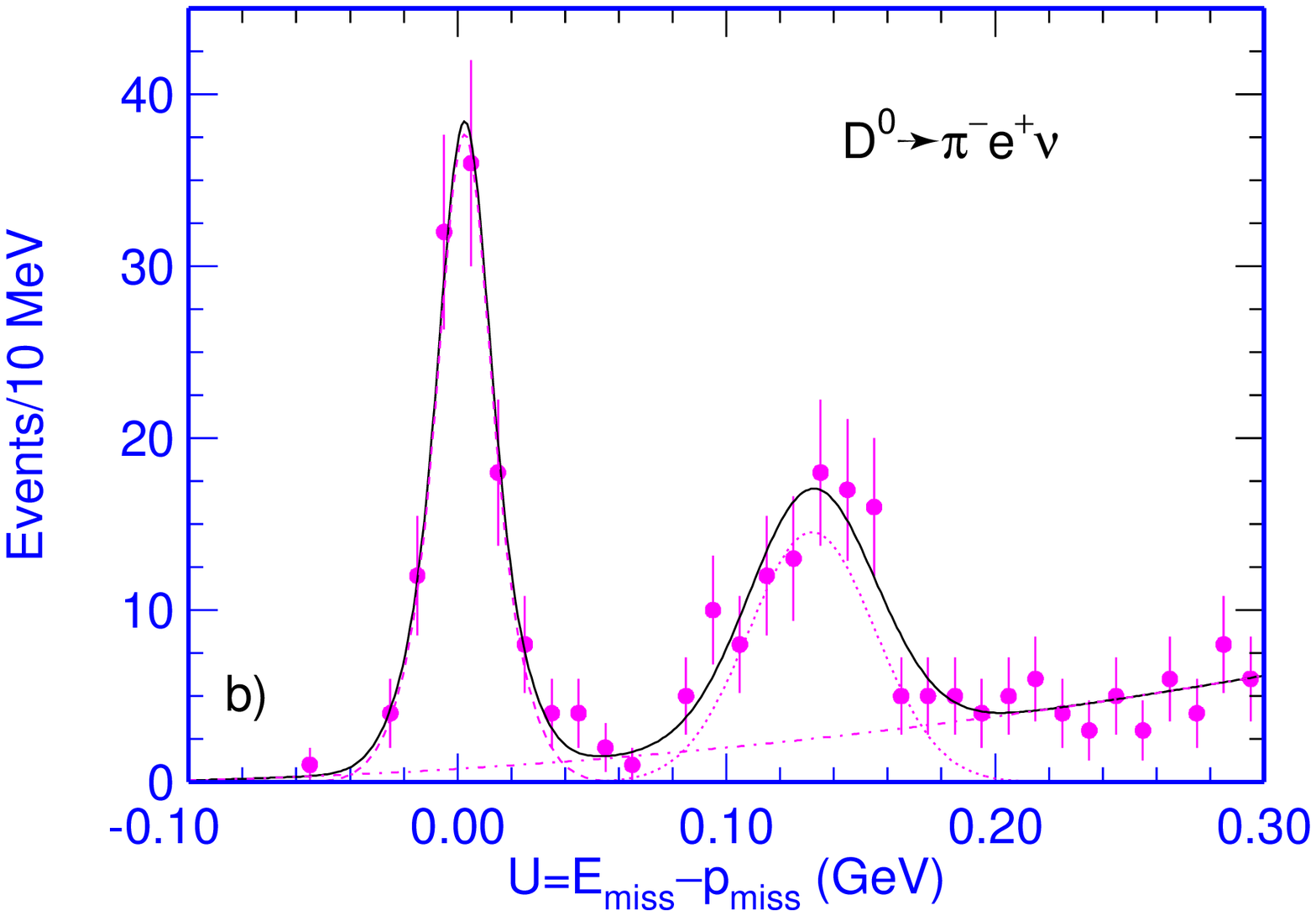}}} 
\centerline{\mbox{\epsfxsize=0.50\textwidth\epsffile{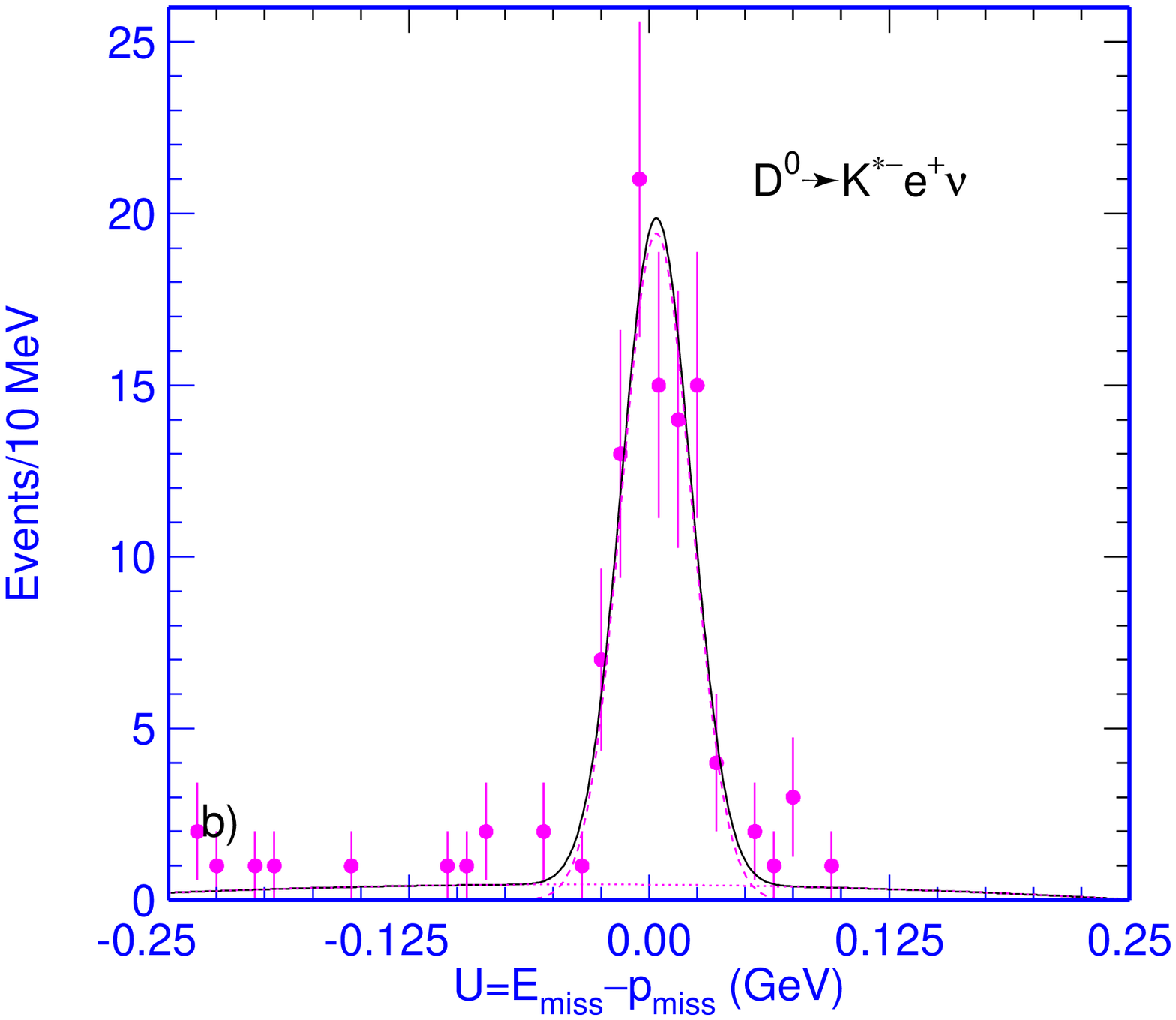}
\epsfxsize=0.50\textwidth\epsffile{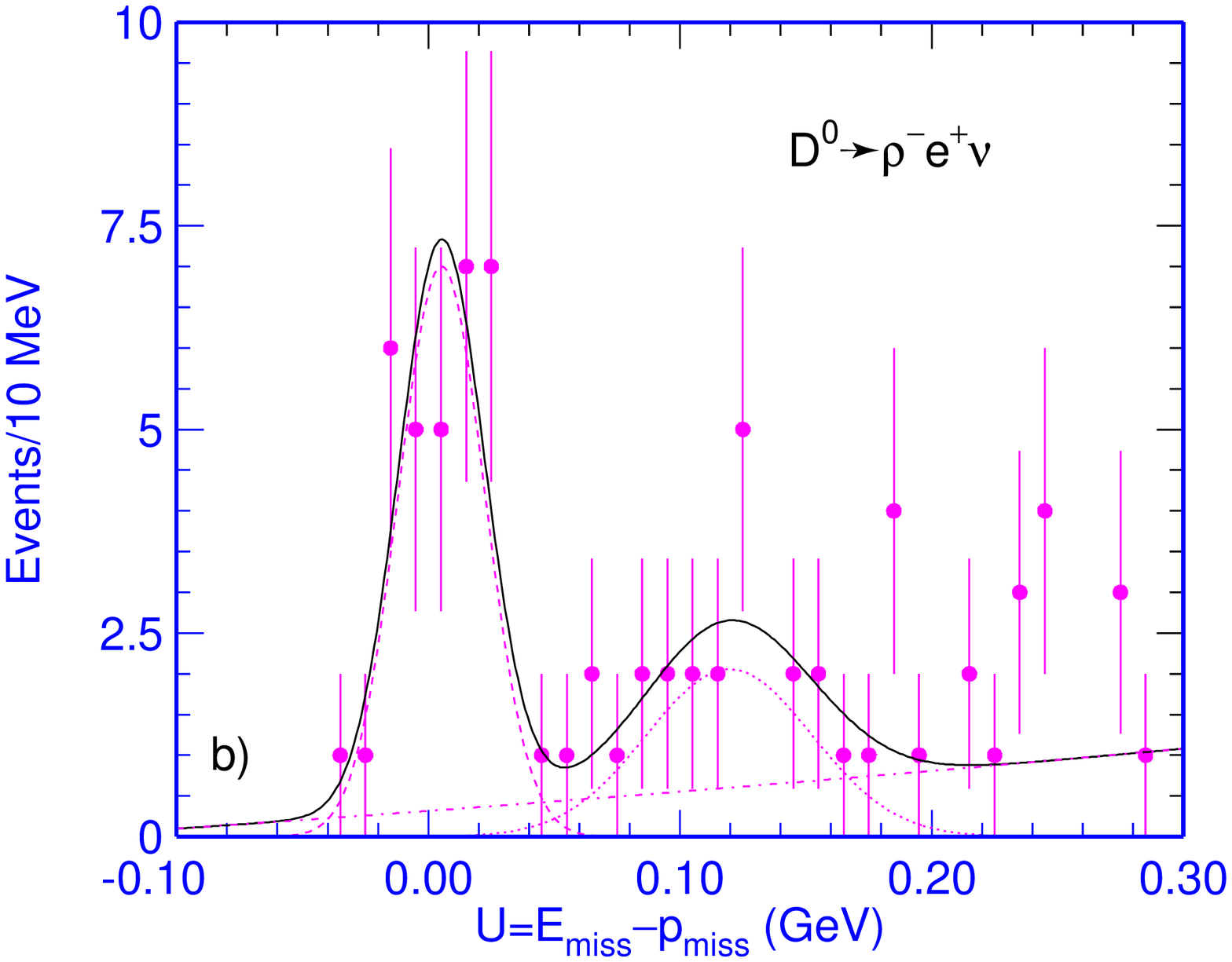}}} 
\caption{Fits to $U=E_{\rm miss}-p_{\rm miss}$ distributions for 
         $D^0 \to \ke$,  $\pie$, $\kste$ and $\rhoe$, with the other
         $\bar{D^{0}}$ fully reconstructed.}
\label{signal-yield}
\end{figure}

The comparison of the $U$ distribution for the data and MC for 
$\drhoe$ is shown in Fig.~\ref{urho}. The background 
decomposition is shown in Fig.~\ref{urho}. 
We fit the $U$ distribution with a Gaussian which corresponds to the
$\drhoe$ signal, another Gaussian for backgrounds mainly from $\dkste$,
and a 2nd order polynomial function determined from MC simulation.
The fit to the data is shown in Fig.~\ref{signal-yield}.
The yields and efficiencies are given in Table~\ref{table2}.  

To determine $\vcs$ and $\vcd$, we need to study the 
$q^2$-dependent differential decay rate. 
Energy-momentum conservation is used to determine $q^2$ in the 
laboratory frame 
\begin{eqnarray}
E_W&=&E_{b}-E_P, \non\\ 
\vec{p}_W&=&-\vec{p}_{D}-\vec{p}_P,\non\\ 
q^2&=&E_W^2-|\vec{p_W}|^2.   
\end{eqnarray} 
Here $E_W$ and $\vec p_W$ are the 
energy  and momentum vector of the $\ell \nu$ system or 
the virtual $W$.  
$\vec{p}_{D}$ is the momentum vector of the tagging $D$ meson.
$E_P$ and $\vec{p}_P$ are the energy and momentum vector of the 
hadronic system in the other $D$ semileptonic decays.

\begin{figure}[htbp]
\epsfxsize=0.33\textwidth\centerline{\mbox{\epsffile{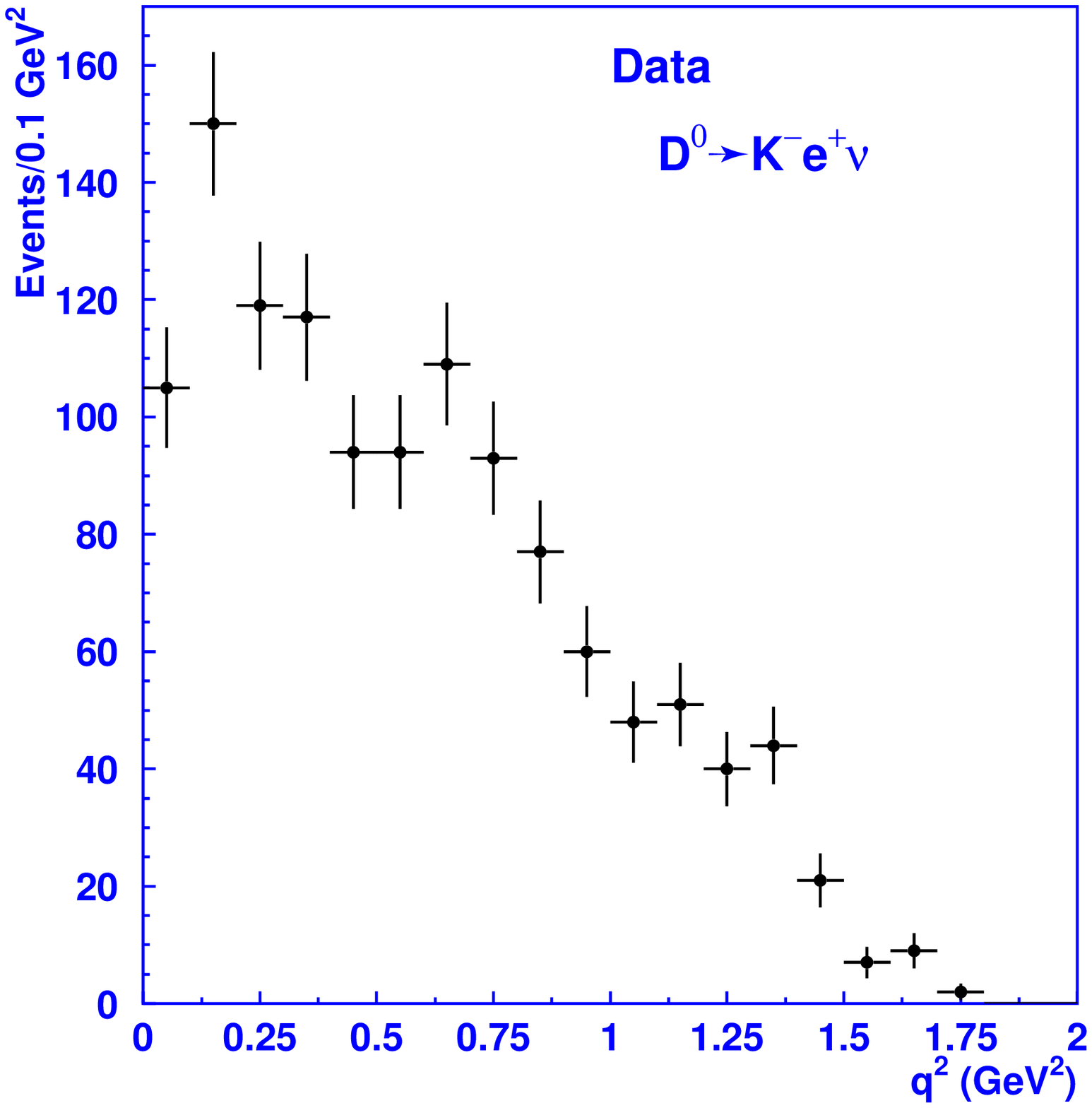}
\epsfxsize=0.33\textwidth\epsffile{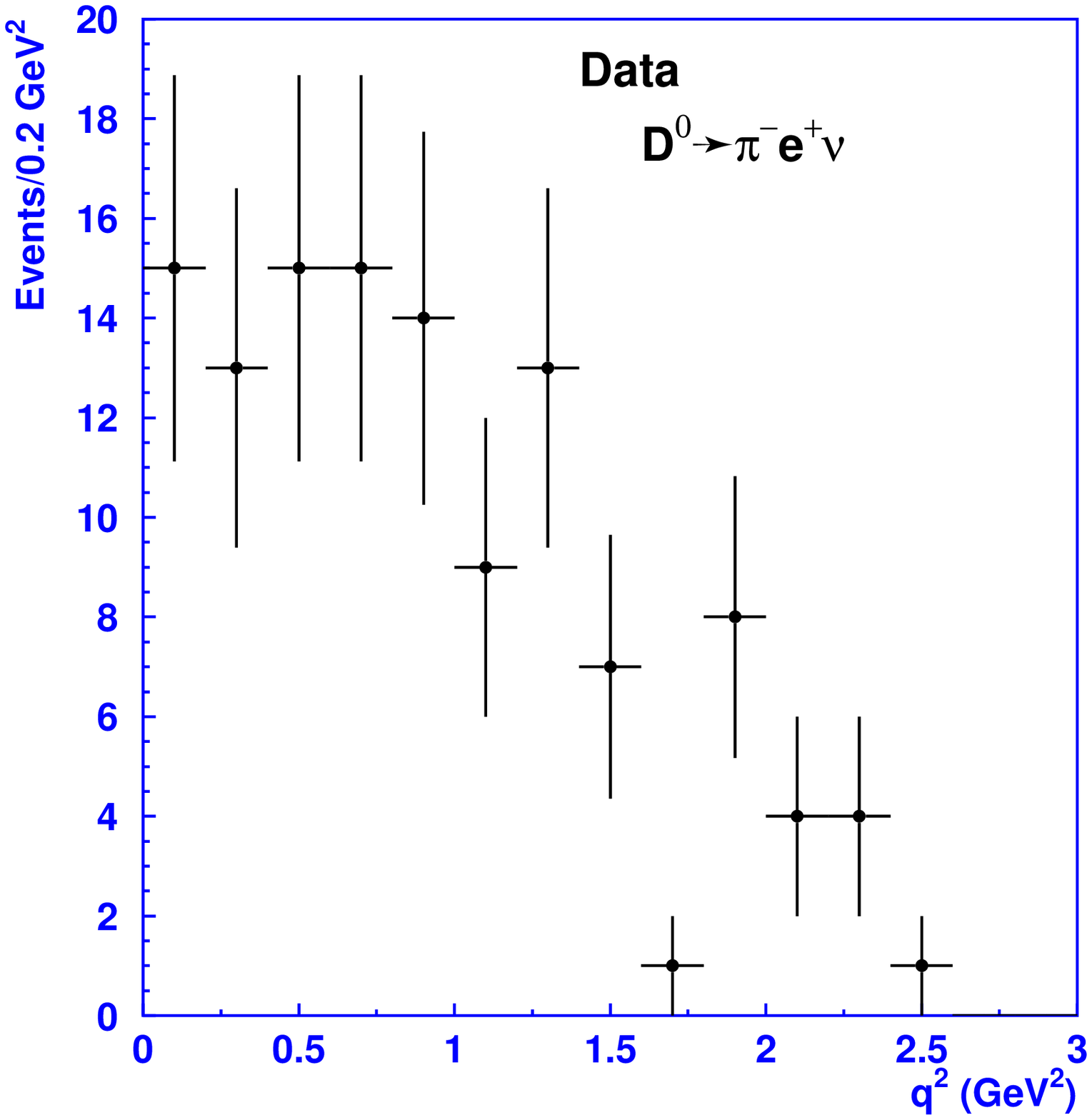}
\epsfxsize=0.33\textwidth\epsffile{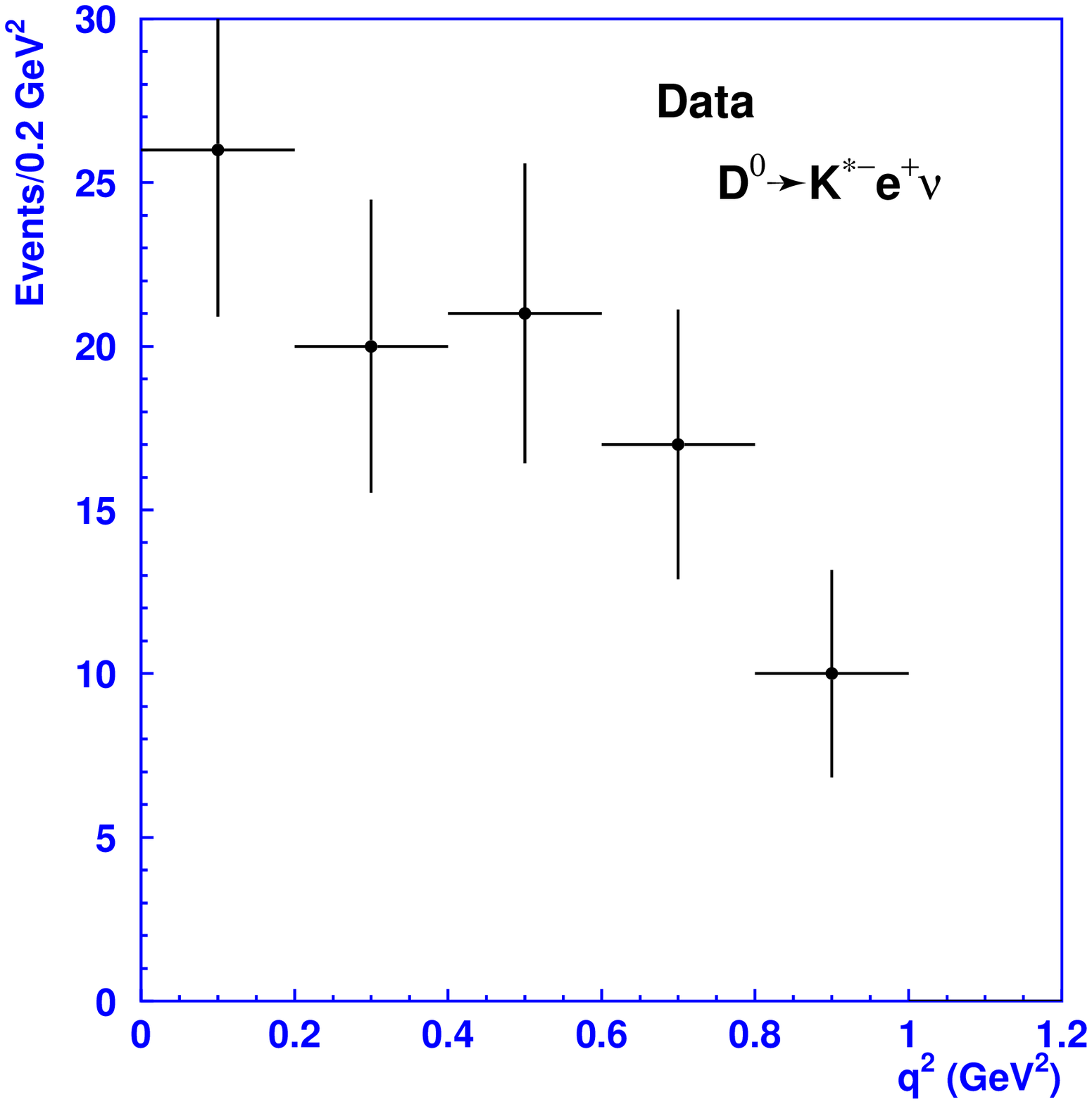}}} 
\caption{Uncorrected $q^2$ distributions for $\dke$, $\dpie$, 
         $\dkste$. }
\label{q23}
\end{figure}

In Fig.~\ref{q23}, we present $q^2$ distributions for $\dke$, 
$\dpie$ and $\dkste$ from data. 
These $q^2$ distributions
are from data directly, without any efficiency correction.
Our $q^2$ resolution is about 0.025 GeV$^{2}$ or smaller which 
is over a factor of 10 better than CLEO III which achieved a 
resolution of 0.4 GeV$^{2}$~\cite{Hsu}. 
This huge improvement in resolution is due to the unique 
kinematics at the $\psi(3770)$ resonance.
The $q^2$ distribution for
$\drhoe$ from data is not shown due to the very limited statistics. 
Detailed study and measurements of form factor and CKM matrix
elements will be performed with more CLEO-c data.

\section{SYSTEMATIC UNCERTAINTIES} 

We have considered the following sources of systematic errors which 
may affect our results: charged track and $\pi^0$ finding, 
the veto on extra tracks, background normalization 
and shape, MC statistics, incomplete modeling of radiative 
corrections, model-dependent form factors, 
hadron identification, and electron identification. 
The contributions of these systematic errors are discussed below.  
 
From studies of CLEO-c data and MC, the systematic error for track 
finding efficiency is estimated to be 3\% per charged track, 
4.4\% per $\pi^0$ and 3\% per $K^{0}_{S}$.
The systematic errors in particle identification are
2\% for electron and 1\% for pion or kaon.
MC statistics varies from 1\% to 3\%, depending on the tag modes. 
To test the potential effect of uncertainties in radiative 
corrections, we have generated signal MC with and without ISR, 
and have found that the efficiencies vary within 1\%. 
Using the selected $\dke$ events, we find 0.51\% events have extra 
tracks in our data, 0.41\% events in our MC. Therefore, 
we assign a 0.5\% systematics uncertainty due to the veto on 
extra tracks.  
To estimate the systematics from the backgrounds, we vary the 
background functions by $\pm1\sigma$, taking the correlation of 
parameters into account. We find the background normalization 
and shape contributes a systematic error of 1.1\%, 3.1\%, 2.9\% and 
5.3\% for $\dke$, $\dpie$, $\dkste$ and $\drhoe$, respectively.  
Systematic uncertainties on the reconstruction efficiencies
from our current knowledge of form factors are small ($<$ 2\%) 
due to the uniform acceptance of the CLEO-c detector over large 
solid angle and momentum range.

Most of these systematic errors are estimated from initial study of 
the first CLEO-c data. We expect these systematic errors to be 
reduced in the future, with more detailed study and more CLEO-c data.

In Table~\ref{systematics}, we summarize the contributions from
systematic errors discussed above.  The systematic errors from MC 
statistics is weighted with tag yields. We add these sources in 
quadrature to obtain the total systematic error for each decay mode.

\begin{table}[hbtp]
\caption{Systematics for four $D^0$ semileptonic decays (in \%).}
\begin{center}
\begin{tabular}{lccccc}\hline
sources         & $\dke$ & $\dpie$ & $\dkste$ & $\drhoe$ \\ \hline 
tracking        & 6    & 6     & 6      & 6    \\ 
$\pi^0$ finding & -    & -     & 4.4    & 4.4    \\ 
EID             & 2    & 2     & 2      & 2    \\ 
PID             & 1    & 1     & 1      & 1    \\ 
extra track     & 0.5  & 0.5   & 0.5    & 0.5  \\ 
MC statistics   & $<$1 & 1.1   & 2.2    & 1.9  \\ 
backgrounds     & 1.1  & 3.1   & 2.9    & 5.3  \\ 
ISR             & 1    & 1     & 1      & 1    \\ 
Form factors    & $<2$ & $<2$  & $<2$   & 5.0  \\  
Yields          &   1  & 1.9   & 1      & 2.7  \\ \hline 
Total           & 7.0  & 7.8   & 8.9    & 11.2 \\ \hline 
\end{tabular}
\end{center} 
\label{systematics}
\end{table}

\section{SUMMARY}

In summary, we present improved measurements of exclusive $D^0$ 
semileptonic decays using the first CLEO-c data. The absolute 
branching fractions are listed in Table~\ref{summary}.

\begin{table}[htbp]
\caption{Absolute branching fraction measurements of the exclusive 
         $D^0$ semileptonic decays, in comparison with PDG~\cite{PDG}.
         The uncertaities are statistical and systematic, respectively.}  
\begin{center}
\begin{tabular}{lccccc}\hline
 Decays          & ${\cal B}$  & PDG  \\ \hline 
$\dke$     & (3.52$\pm0.10\pm0.25)\%$   & $(3.58\pm0.18)\%$ \\ 
$\dpie$    & (0.25$\pm0.03\pm0.02)\%$   & $(0.36\pm0.06)\%$ \\ 
$\dkste$   & (2.07$\pm0.23\pm0.18)\%$   & $(2.15\pm0.35)\%$ \\ 
$\drhoe$   & (0.19$\pm0.04\pm0.02)\%$   & none \\  \hline 
{${\cal B}(\dpie)\over{\cal B}(\dke)$}  & $(7.0\pm0.7\pm0.3)\%$ 
           & $(10.1\pm1.8)\%$ \\ 
{${\cal B}(\drhoe)\over{\cal B}(\dkste)$} & $(9.2\pm2.0\pm0.8)\%$ 
           & none \\ \hline 

\end{tabular}
\end{center}
\label{summary}
\end{table}

Our results for $\dke$ and $\dkste$ are consistent with those from 
the PDG~\cite{PDG}. Our ${\cal B}(\dpie)$ result is lower than 
that from the PDG~\cite{PDG}. 
The ratio {${\cal B}(\dpie)\over{\cal B}(\dke)$} is closer to the 
CLEO III result~\cite{Hsu} of $(8.2\pm0.6\pm0.5)\%$, while lower than 
the PDG value~\cite{PDG}. 
The improved precision in these branching fractions is consistent
with the expected performance of CLEO-c with initial data sample
of 60 pb$^{-1}$ at the $\psi(3770)$. The full integrated luminosity
of $\sim$3 fb$^{-1}$ can be expected to provide great improvement
in our knowledge of a much larger set of $D^{0}$ and $D^{+}$
exclusive semileptonic branching fractions as well as decay form
factors and the CKM matrix elements $\vcs$ and $\vcd$~\cite{cleoc}.

We gratefully acknowledge the effort of the CESR staff in providing 
us with excellent luminosity and running conditions.
This work was supported by the National Science Foundation,
the U.S. Department of Energy, the Research Corporation,
and the Texas Advanced Research Program.

\end{document}